%% file: main.tex
\documentclass[12pt]{ociamthesis}  

\usepackage{amssymb}
\usepackage{mathtools} 
\allowdisplaybreaks
\usepackage{subcaption}
\usepackage{graphicx}
\usepackage{tikzit}
\usepackage{xcolor}
\usepackage{listings}
\usepackage{xparse}
\usepackage{kantlipsum}
\usepackage{algorithm}
\usepackage{algorithmicx}
\usepackage[noend]{algpseudocode}
\usepackage{tikz}
\usetikzlibrary{arrows.meta, chains, positioning, shapes.geometric, decorations.pathreplacing}
\usepackage{siunitx}
\usepackage{booktabs}
\usepackage{multirow}
\usepackage[stable]{footmisc}

\tikzset{FlowChart/.style={
    startstop/.style = {rectangle, draw, fill=red!30, minimum width=3cm, minimum height=1cm,
      on chain, join=by arrow},
    process/.style = {rectangle, rounded corners, draw, fill=blue!30, text width=5cm,
      minimum height=1cm, align=center, on chain, join=by arrow},
    decision/.style = {diamond, aspect=1.3, draw, fill=green!30, minimum width=3cm,
      minimum height=1cm, align=center, on chain, join=by arrow},
    arrow/.style = {thick,-Triangle}
  }
}

\input{zx.tikzstyles}
\input{zx.tikzdefs}

\newcommand{\bra}[1]{\ensuremath{\left\langle #1 \right|}}
\newcommand{\ket}[1]{\ensuremath{\left|  #1 \right\rangle}}

\newcommand{\ketbra}[2]{\ensuremath{\ket{#1}\!\bra{#2}}}

\newcommand*\Let[2]{\State #1 $\gets$ #2}
\algrenewcommand\algorithmicrequire{\textbf{Precondition:}}
\algrenewcommand\algorithmicensure{\textbf{Postcondition:}}

\NewDocumentCommand{\codeword}{v}{
  \texttt{\textcolor{gray}{#1}}
}
\title{Vanishing 2-Qubit Gates with\\[1ex]     
        Non-Simplification ZX-Rules}   

\author{Ryan Krueger}            
\college{Christ Church College}  

\degree{Master of Science in Mathematical Sciences}     
\degreedate{Trinity 2021}         

\begin{document}

\baselineskip=18pt plus1pt

\setcounter{secnumdepth}{3}
\setcounter{tocdepth}{3}

\maketitle                  
\include{frontmatter/acknowledgements}  
\include{frontmatter/abstract}          

\begin{romanpages}          
\tableofcontents            
\end{romanpages}            

\include{chapters/introduction}
\include{chapters/background}
\include{chapters/methods}
\include{chapters/results}
\include{chapters/discuss-conc}


\addcontentsline{toc}{chapter}{Bibliography}
\bibliography{refs}        
\bibliographystyle{plain}  

\end{document}

%% file: zx.tikzstyles
\tikzstyle{box}=[shape=rectangle, text height=1.5ex, text depth=0.25ex, yshift=0.5mm, fill=white, draw=black, minimum height=5mm, yshift=-0.5mm, minimum width=5mm, font={\small}]
\tikzstyle{Z dot}=[inner sep=0mm, minimum size=2mm, shape=circle, draw=black, fill={rgb,255: red,221; green,255; blue,221}]
\tikzstyle{Z phase dot}=[minimum size=5mm, font={\footnotesize\boldmath}, shape=rectangle, rounded corners=2mm, inner sep=0.2mm, outer sep=-2mm, scale=0.8, tikzit shape=circle, draw=black, fill={rgb,255: red,221; green,255; blue,221}, tikzit draw=blue]
\tikzstyle{X dot}=[Z dot, shape=circle, draw=black, fill={rgb,255: red,255; green,136; blue,136}]
\tikzstyle{X phase dot}=[Z phase dot, tikzit shape=circle, tikzit draw=blue, fill={rgb,255: red,255; green,136; blue,136}, font={\footnotesize\boldmath}]
\tikzstyle{hadamard}=[fill=yellow, draw=black, shape=rectangle, inner sep=0.6mm, minimum height=1.5mm, minimum width=1.5mm]
\tikzstyle{vertex}=[inner sep=0mm, minimum size=1mm, shape=circle, draw=black, fill=black]
\tikzstyle{vertex set}=[inner sep=0mm, minimum size=1mm, shape=circle, draw=black, fill=white, font={\footnotesize\boldmath}]

\tikzstyle{hadamard edge}=[-, dashed, dash pattern=on 2pt off 0.5pt, thick, draw={rgb,255: red,68; green,136; blue,255}]
\tikzstyle{brace edge}=[-, tikzit draw=blue, decorate, decoration={brace,amplitude=1mm,raise=-1mm}]
\tikzstyle{diredge}=[->]

%% file: frontmatter/acknowledgements.tex
\begin{acknowledgements}
  I would like to thank my supervisor Aleks Kissinger for coordinating this project with me and for his mentorship throughout the year.
  I would also like to thank my family for their love and support.
\end{acknowledgements}

%% file: frontmatter/abstract.tex
\begin{abstract}

Traditional quantum circuit optimization is performed directly at the circuit level.
Alternatively, a quantum circuit can be translated to a ZX-diagram which can be simplified using the rules of the ZX-calculus, after which a simplified circuit can be extracted.
However, the best-known extraction procedures can drastically increase the number of 2-qubit gates.
In this work, we take advantage of the fact that local changes in a ZX-diagram can drastically affect the complexity of the extracted circuit.
We use a pair of congruences (i.e., non-simplification rewrite rules) based on the graph-theoretic notions of local complementation and pivoting to generate local variants of a simplified ZX-diagram.
We explore the space of equivalent ZX-diagrams generated by these congruences using simulated annealing and genetic algorithms to obtain a simplified circuit with fewer 2-qubit gates.
On randomly generated circuits, our method can outperform state-of-the-art optimization techniques for low-qubit ($<10$) circuits.
On a set of previously reported benchmark circuits with $\leq 14$ qubits, our method outperforms off-the-shelf methods in 87\% of cases, consistently reducing overall circuit complexity by an additional \textasciitilde 15-30\% and eliminating up to 46\% of 2-qubit gates.
These preliminary results serve as a proof-of-concept for a new circuit optimization strategy in the ZX-calculus.

\end{abstract}

%% file: chapters/introduction.tex
\chapter[Introduction]{Introduction} \label{ch:intro}


Quantum computing arose from the idea that the ``bugs'' of quantum mechanics (e.g., superposition) may serve as useful features for information processing.
Since this framing was posed in the late 1970's, a rich theory of quantum information has developed that treats a \emph{qubit}, a superposition of two classical bits, as the basic unit of information.
Quantum information science is an exciting field of work as many quantum algorithms have been discovered that offer significant speedups from their classical counterparts (e.g., Shor's algorithm for integer factorization).
However, like the formulation of the Turing machine decades before the invention of the transistor, efforts to engineer a real-world quantum computer that can implement these theories are in their infancy.

The current state-of-the-art are so-called noisy intermediate-scale quantum (NISQ) computers~\cite{preskill2018quantum}.
These are devices without enough qubits to spare for error correction (i.e., noisy) and with a relatively small qubit number (i.e., intermediate-scale).
With \textasciitilde 50 qubits, NISQ devices cannot run particularly resource-intensive quantum algorithms (e.g., Shor's algorithm at a meaningful scale) but do afford the occasional speedup over classical computers.
To stress these capabilities, there is significant effort towards minimizing \emph{quantum circuits}, the standard model for a quantum computation, to mitigate noise and decoherence.
More formally, this field is known as \emph{quantum circuit optimization}.
The ZX-calculus, a graphical formalism for quantum circuits rooted in category theory, provides a unique setting for circuit optimization.
In this language, quantum circuits are represented as \emph{ZX-diagrams} that are manipulated via a set of rewrite rules.
ZX-diagrams provide a lower-level representation for quantum computations and can be deformed arbitrarily, enriching the space of possible simplifications.
While there is no known procedure for recovering a quantum circuit from a general ZX-diagram, subsets of transformations have been identified that preserve circuit extractability.
These transformations have enabled a suite of circuit optimization procedures in the ZX-calculus (e.g., for reducing T-count) with various tradeoffs.

Still, circuit extraction from a ZX-diagram remains a bottleneck for optimization;
the best known extraction procedures can introduce an unwieldy degree of complexity in the resulting circuit.
For example, circuit extraction can introduce many CNOT gates which are particularly problematic on NISQ devices.
Interestingly, however, closely related ZX-diagrams can induce drastically different circuits.
Current optimization procedures employing the ZX-calculus do not exploit this fact (e.g., by searching local variants) but rather return a ZX-diagram (or its associated circuit) once no more graph simplifications can be applied.

In this thesis, we explore the role that such a search over local variants of ZX-diagrams could play in circuit optimization.
More specifically, we apply various search procedures (i.e., simulated annealing and genetic algorithms) over the application of congruences (i.e., non-simplification rewrite rules) to a fully simplified ZX-diagram.
We use two congruences that arise from the graph-theoretic notions of local complementation and pivoting.
We find that our optimization strategy outperforms off-the-shelf methods on randomly generated circuits with $<10$ qubits and on a set of benchmark circuits with $\leq 14$ qubits.
Most notably, on the benchmark circuits, our method eliminates up to 46\% of 2-qubit gates and consistently reduces the circuit complexity by an additional 15-30\%.

%% file: chapters/background.tex
\chapter[Background]{Background} \label{ch:bg}


To understand quantum circuit optimization in the ZX-calculus, one first must understand quantum circuits and the ZX-calculus.
In this section we provide the requisite background in both of these, preceded by a brief discussion on qubits.
We conclude with an overview of quantum circuit optimization in the ZX-calculus.

\section{Qubits}\label{sec:qubits}

A classical bit has a value of 0 or 1.
A quantum bit, or \emph{qubit}, encodes a quantum superposition of these two values and therefore more information than a classical bit.
In the traditional bra-ket notation of quantum mechanics, a qubit is represented by a vector in $\mathbb{C}^2$,
\begin{align*}
  |\psi\rangle = \lambda_1 |0\rangle + \lambda_2 |1\rangle = \begin{pmatrix}\lambda_1 \\ \lambda_2 \end{pmatrix}
\end{align*}
where
\begin{align*}
  & |0\rangle = \begin{pmatrix}1 \\ 0\end{pmatrix} \\
  & |1\rangle = \begin{pmatrix}0 \\ 1\end{pmatrix}
\end{align*}
and $\lambda_1$ and $\lambda_2$ are the probability amplitudes of observing $|0\rangle$ and $|1\rangle$ upon measurement in this basis, respectively.
Notably, qubits can be written with respect to any orthonormal basis.
The set $\{|0\rangle, |1\rangle\}$ is known as the \emph{computational} basis.
Another common basis is the +/- basis, where
\begin{align*}
  & |+\rangle = \frac{|0\rangle + |1\rangle}{\sqrt{2}} \\
  & |-\rangle = \frac{|0\rangle - |1\rangle}{\sqrt{2}}
\end{align*}
The computational and +/- bases are also known as the Z and X bases, respectively.

More generally, a system of $n$ qubits is represented by the tensor product of the individual states.
So, an $n$-qubit state corresponds to a $2^{n}$-dimensional vector.
For example, the two-qubit state with both qubits in $|0\rangle$ is
\begin{align*}
  |0\rangle \otimes |0\rangle = |00\rangle = \begin{pmatrix}1 \\ 0 \\ 0 \\ 0\end{pmatrix}
\end{align*}
In the ZX-calculus and this work, qubits are abstracted as wires in string diagrams.
However, this definition is included both for completeness and to aid understanding of quantum circuits.

\begin{figure}
\centering
\begin{subfigure}{.5\textwidth}
  \centering
  \includegraphics[width=.4\linewidth]{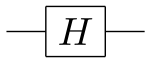}
  \caption{The Hadamard gate.}
  \label{fig:had}
\end{subfigure}%
\begin{subfigure}{.5\textwidth}
  \centering
  \includegraphics[width=.4\linewidth]{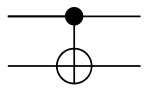}
  \caption{The CNOT gate. The top and bottom qubits are the control and target, respectively.}
  \label{fig:cnot}
\end{subfigure}
\caption{
  The circuit representation of two common quantum logic gates.
  Wires represent qubits and boxes represent quantum logic gates.
  Quantum circuits are read from left to right.
}
\label{fig:test}
\end{figure}

\section{Quantum Circuits}\label{sec:qcircs}

A quantum circuit models a quantum computation as a sequence of discrete gates.
In the traditional notation with qubits as vectors in $\mathbb{C}^2$, quantum gates correspond to unitary matrices.
For example, the Hadamard gate, a quantum gate that acts on a single qubit and maps
\begin{align*}
  & |0\rangle \mapsto \frac{|0\rangle + |1\rangle}{\sqrt{2}} \\
  & |1\rangle \mapsto \frac{|0\rangle - |1\rangle}{\sqrt{2}}
\end{align*}
has the matrix form
\begin{align*}
  H = \frac{1}{\sqrt{2}}\begin{bmatrix}1 & 1 \\ 1 & -1\end{bmatrix}
\end{align*}
Figure \ref{fig:had} depicts the circuit representation of the Hadamard gate.
Note that $|+\rangle = H|0\rangle$ and $|-\rangle = H|1\rangle$.
An important family of single-qubit gates are the \emph{phase shift} gates that modify the phase of the quantum state by some $\alpha$.
A phase shift gate maps
\begin{align*}
  & |0\rangle \mapsto |0\rangle \\
  & |1\rangle \mapsto e^{i\alpha}|1\rangle \\
\end{align*}
and has the matrix form
\begin{align*}
  R_{\alpha} = \begin{bmatrix}1 & 0 \\ 0 & e^{i\alpha}\end{bmatrix}
\end{align*}
Examples of phase gates are the T-gate ($\alpha = \frac{\pi}{4}$) and the S-gate ($\alpha = \frac{\pi}{2}$).
Phase gates where $\alpha$ is a multiple of $\pi / 2$ are called \emph{Clifford} gates.

Quantum gates are not restricted to acting on single qubits.
For example, the controlled-NOT (CNOT) gate acts on two qubits and flips the second (target) qubit only when the first (control) qubit is $|1\rangle$.
The CNOT gate is typically depicted as shown in Figure \ref{fig:cnot} and has the following matrix form:
\begin{align*}
  CNOT = \begin{bmatrix}1 & 0 & 0 & 0 \\0 & 1 & 0 & 0 \\ 0 & 0 & 0 & 1\\0 & 0 & 1 & 0\end{bmatrix}
\end{align*}
In general, an $n$-qubit gate is represented by a $2^n$-dimensional unitary.
However, attention is typically restricted to single- and 2-qubit gates as it is well known that the Clifford + T gate set, consisting of the H, T, S, and CNOT gates, is \emph{universal} -- any other operation can be represented by a finite sequence from this set.
The \emph{Clifford circuits} are those that can be generated by $\{H, S, CNOT\}$.
As the T and S gates are instances of phase shift gates, $\{H, R_{\alpha}, CNOT\}$ is also universal though in practice any non-Clifford phase gate is implemented with T gates. 

Quantum gates can be composed sequentially or in parallel.
Consider two gates $A$ and $B$.
The effect of $B$ applied in series after $A$ can be described by a single gate (i.e., linear map) $B \cdot A$ via matrix multiplication.
Alternatively, the tensor product $B \otimes A$ is used to describe A and B in parallel.
For example, the simple circuit shown in Figure \ref{fig:simple-trad} can be described by the following linear map:
\begin{align*}
  (\mathbb{I} \otimes CNOT) \cdot (\mathbb{I} \otimes H \otimes \mathbb{I}) \cdot (CNOT \otimes \mathbb{I}) \cdot (S \otimes \mathbb{I} \otimes \mathbb{I})
\end{align*}

\begin{figure}
\centering
\input{ZX-rules.tikz}
\caption{
  The rules of the ZX-calculus.
  All rules hold with the colors interchanged and for $\alpha, \beta \in [0, 2 \pi)$.}
\label{fig:zx-rules}
\end{figure}
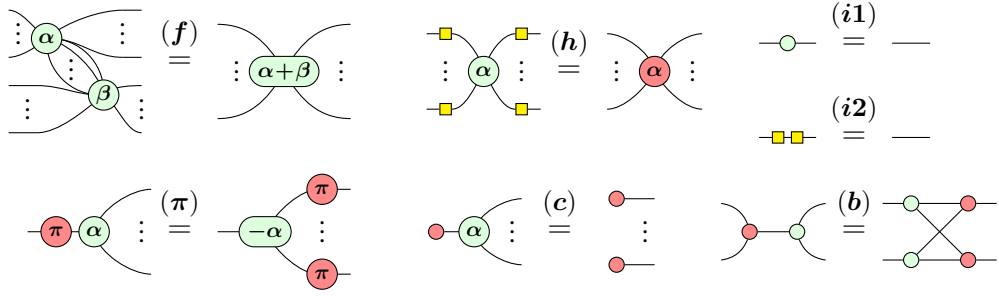

\section{ZX-Calculus\footnote{Many of the diagrams used in this section were originally presented in \cite{duncan2020graph}}}\label{sec:zx}

The ZX-calculus is a graphical language for representing and reasoning about quantum processes.
Quantum processes are represented by \emph{ZX-diagrams} which consist of \emph{wires} and \emph{spiders}.
Spiders can have an arbitrary number of inputs and outputs and come in two flavors: Z spiders (shown in green) and X spiders (shown in red).
As with quantum circuits, a ZX-diagram is interpreted from left to right.

ZX-diagrams abstract away the complexities of linear algebra and tensor products from traditional formalisms of quantum processes.
Wires still represent qubits while spiders provide a general form for linear maps, where
\begin{align*}
  & \input{Zsp-a.tikz} := \ketbra{0...0}{0...0} + e^{i\alpha}\ketbra{1...1}{1...1} \\
  & \input{Xsp-a.tikz} := \ketbra{+...+}{+...+} + e^{i\alpha}\ketbra{-...-}{-...-}
\end{align*}
Two diagrams can be composed in serial by joining the outputs of the first to the inputs of the second, or in parallel by stacking the two diagrams.

From this definition, we can identify the diagrammatic represenations for several common components of quantum computation:
\[
\begin{array}{rclcrcl}
\input{ket-+.tikz} & = & \ket{0} + \ket{1} \ \propto \ket{+} &
\qquad\qquad &
\input{ket-0.tikz} & = & \ket{+} + \ket{-} \ \propto \ket{0} \\
&\quad& & & \quad \\
\input{Z.tikz} & = & \ketbra{0}{0} - \ketbra{1}{1} = Z &
&
\input{X.tikz} & = & \ketbra{+}{+} - \ketbra{-}{-} = X
\end{array}
\]
where $Z$ and $X$ are the corresponding Pauli matrices.
Note that spiders without any inputs can be regarded as qubit state preparations.
The Hadamard gate is so pervasive that it merits shorthand notation;
we use a yellow square to represent the Hadamard gate as shown below
\begin{equation}\label{eq:had-short}
  \input{had-alt.tikz}
\end{equation}
and replace a Hadamard gate between two spiders with a blue dashed edge:
\ctikzfig{blue-edge-def}

We can also reason about ZX-diagrams.
The first rule for determining equality is that \emph{only connectivity matters} (OCM).
In other words, two ZX-diagrams are equal when one can be deformed into the other (e.g., via moving vertices around in the plane or bending wires) while maintaining connectivity and the order of the inputs and outputs.
In addition to this OCM principle, the ZX-calculus includes a primary set of rewrite rules shown in Figure \ref{fig:zx-rules}.
These rules only hold up to non-zero scalar factors, though these scalars are typically ignored as they correspond to negligible differences in global phase.
Additional rules can be derived from this set, such as the antipode rule
\ctikzfig{hopf-rule}
and the $\pi$-copy rule
\ctikzfig{picopy-rule}

\begin{figure}
\centering
\begin{subfigure}[t]{.3\textwidth}
  \centering
  \includegraphics[width=4cm]{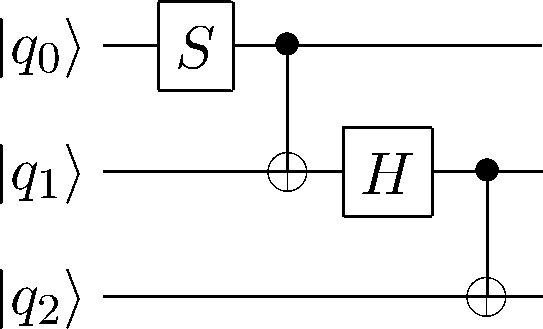}
  \caption{Traditional notation}
  \label{fig:simple-trad}
\end{subfigure}%
\begin{subfigure}[t]{.7\textwidth}
  \centering
  \resizebox{9cm}{!}{\input{figures/zx_full.tikz}}
  \caption{ZX-diagram}
  \label{fig:simple-zx}
\end{subfigure}
\caption{A simple 3-qubit quantum circuit shown in both the traditional notation and as a ZX-diagram.}
\label{fig:simple-circ}
\end{figure}

All quantum circuits can be translated to ZX-diagrams.
This is a consequence of the fact that the following universal gate set can be easily represented in the ZX-calculus:
\begin{align*}
CNOT & = \input{cnot.tikz} &
R_{\alpha} & = \input{Z-a.tikz} &
H & = \input{h-alone.tikz}
\end{align*}
In the CNOT diagram, the green and red spiders are the control and target qubits, respectively.
As an example, Figure \ref{fig:simple-circ} depicts both the traditional and diagrammatic representations of a simple quantum circuit.

\section{Circuit Optimization in the ZX-Calculus}\label{sec:zx-circ-opt}


The goal of quantum circuit optimization is to simplify quantum circuits (e.g., via reducing the number of gates) to reduce noise and decoherence when run on modern quantum computers.
More specifically, the T-count and 2-qubit-count are typically targeted; fault tolerant implementations of T gates generally require an order of magnitude more resources than Clifford gates~\cite{campbell2017roads}, and the fidelity of single-qubit gates is demonstrably higher than that of 2-qubit gates~\cite{ballance2016high}.
It is standard to perform optimizations at the circuit-level.
Examples of circuit-level optimizations are gate decompositions~\cite{vartiainen2004efficient} and cancelling back-to-back CNOT gates~\cite{garcia2011equivalent}.

The ZX-calculus enables circuit optimization at the graph-level, a more flexible setting that is well-studied in its own right.
Such an optimization involves converting a circuit to a ZX-diagram, simplifying the diagram, and converting the diagram back to a circuit. 
However, there is no known general-purpose procedure for recovering a quantum circuit from a generic ZX-diagram.
This restricts the space of permissible simplifications to those that preserve whichever diagrammatic properties are required by the extraction procedure.

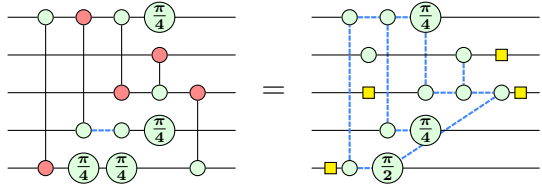
\begin{figure}
\centering
\input{graph-like-ex.tikz}
\caption{An example of a ZX-diagram and an equivalent, graph-like ZX-diagram.}
\label{fig:graph-like}
\end{figure}

In 2020, Duncan et al. reported an extraction procedure that enabled current best practices~\cite{duncan2020graph}.
Firstly, this procedure operates on \emph{graph-like} ZX-diagrams.
A ZX-diagram is graph-like when:
\begin{enumerate}
\item
  All spiders are Z-spiders.
\item
  Z-spiders are only connected via Hadamard edges
\item
  There are no parallel Hadamard edges or self-loops
\item
  Every input or output is connected to a Z-spider and every Z-spider is connected to at most one input or output
\end{enumerate}
Every ZX-diagram is equal to a graph-like diagram (see Figure \ref{fig:graph-like} for an example), and this form admits an underlying graph structure that permits graph-theoretic analyses.
Secondly, this procedure requires that the underlying graph of the input ZX-diagram satisfies a graph-theoretic invariant called \emph{focused generalized flow} (focused gFlow). 
Given such an input ZX-diagram, extraction proceeds in such a way that each non-zero phase corresponds to one phase gate in the resulting circuit but each edge can correspond to multiple CNOTs.
For this reason, local changes in a ZX-diagram (e.g., connectivity) can have significant effects on the complexity of the associated circuit.
Importantly, this extraction procedure is not optimal;
if a given circuit is converted to a ZX-diagram and the extraction procedure is immediately applied, the extracted circuit is often more complex than the original circuit.
In particular, this procedure can drastically increase the number of 2-qubit gates.
For more details on the extraction procedure or focused gFlow, see \cite{duncan2020graph}.


Alongside this extraction procedure, Duncan et al. also introduced an optimization procedure that relies on focused gFlow-preserving graph simplifications.
More specifically, this procedure relies on two graph-theoretic transformations that each correspond to rewrite rules in the ZX-calculus: \emph{local complementation} (LC) and \emph{pivoting}.
Given a graph $G$ and some vertex $u$ of $G$, the \emph{local complementation} of $G$ according to $u$, denoted $G \star u$, is the graph in which the connectivity of all pairs of neighbors of $u$ is inverted.
For example,
\begin{equation*}
G\quad\input{graph1-lab.tikz}\qquad\qquad G\star a\quad\input{graph1-lab-1.tikz}
\end{equation*}
Given a graph-like ZX-diagram, local complementation can be induced in the underlying graph structure while maintaining equality by applying an $X_{-\pi/2}$ gate on (the spider corresponding to) $u$ and a $Z_{\pi/2}$ gate to its neighbors~\cite{duncan2009graph}:
\begin{equation}\label{eq:gs-local-comp}
  \input{local-comp-ex.tikz}
\end{equation}
Relatedly, given two connected vertices $u$ and $v$ in $G$, the \emph{pivot} of $G$ along the edge $uv$ is the graph $G \wedge uv :=G \star u \star v \star u$.
In practice, this consists of complementing the edges between three subsets of vertices: (A) the common neighborhood of $u$ and $v$, (B) the exclusive neighborhood of $u$, and (C) the exclusive neighborhood of $v$:
\[G \quad\input{pivot-L.tikz}\qquad\qquad \quad G\wedge uv \quad\input{pivot-R.tikz}
\]
Again, given a graph-like ZX-diagram, the rewrite rule reported in \cite{duncan2013pivoting} introduces a pivot by applying Hadamard gates on $u$ and $v$, and $Z_{\pi}$ gates on their common neighborhood:
\begin{equation}\label{eq:gs-pivot}
  \input{pivot-desc.tikz}
\end{equation}
Each rewrite rule can be extended to a simplification by performing local complementation (resp. pivoting), removing the vertex (resp. the pair of vertices), and updating the phases:
\ctikzfig{lc-simp}
\ctikzfig{pivot-simp}
Importantly, the existence of a focused gFlow is preserved in both cases.
At a high-level, optimization can then be performed by applying these simplifications to fixpoint and extracting a circuit from the resulting ZX-diagram.
Details on this optimization procedure and proofs of the simplification rules or their focused gFlow-preservation can be found in \cite{duncan2020graph}.
A variant of this procedure is implemented in the \codeword{full_reduce} method of the PyZX library.

One issue with \codeword{full_reduce} is its reliance on circuit extraction; it is not uncommon that the circuit extracted from the final, simplified ZX-diagram has more gates than the input circuit.
Kissinger and van de Wetering introduced an alternative optimization method that uses \emph{phase teleportation} to sidestep the issue of circuit extraction altogether~\cite{kissinger2019reducing}.
Crucially, simplification of the ZX-diagram is performed symbollically.
This symbolic simplification leverages the fact that several rewrite rules involving \emph{phase-gadgets}, a particular ZX-diagrammatic motif, correspond to changes in the original phases.
Therefore, non-Clifford phases can potentially cancel or combine with each other while leaving the original graphical structure of the ZX-diagram intact.
So, as simplification proceeds, application of these rules is simply tabulated to enable reconstruction of a final circuit with the same connectivity but a possible reduction in non-Clifford gates (and therefore T-count).
Importantly, using phase teleportation rather than circuit extraction ensures that the number of 2-qubit and Hadamard gates is unchanged. 
By changing the angles of many non-Clifford phase gates to either 0 or multiples of $\pi /2$, this procedure can also make a circuit-level optimization procedure much more effective.
This procedure is implemented in the \codeword{teleport_reduce} method of PyZX.

In both cases, it is standard to apply a set of circuit-level optimizations post-simplification.
This set of optimizations is implemented in the \codeword{basic_optimization} routine in PyZX.
Note that the circuit must first be extracted after \codeword{full_reduce} whereas these optimizations can be applied directly after \codeword{teleport_reduce}.
In summary, the two standard circuit optimization pipelines in the ZX-calculus are \codeword{full_reduce} + \codeword{extract_circuit} + \codeword{basic_optimization} and \codeword{teleport_reduce} + \codeword{basic_optimization}.

%% file: figures/ZX-rules.tikz
\begin{tikzpicture}
	\begin{pgfonlayer}{nodelayer}
		\node [style=Z phase dot] (0) at (-1.5, -0.625) {$\beta$};
		\node [style=none] (1) at (-0.5, -0.375) {};
		\node [style=none] (2) at (-3.25, -0.375) {};
		\node [style=none] (3) at (-3.25, -1.625) {};
		\node [style=none] (4) at (-0.5, -1.625) {};
		\node [style=none, rotate=90] (7) at (-3.5, -1) {...};
		\node [style=none, rotate=90] (8) at (-0.5, -1) {...};
		\node [style=Z phase dot] (10) at (-3, 0.875) {$\alpha$};
		\node [style=none] (11) at (-1, 1.625) {};
		\node [style=none] (12) at (-4, 1.625) {};
		\node [style=none] (13) at (-1, 0.375) {};
		\node [style=none, rotate=90] (14) at (-1, 1) {...};
		\node [style=none] (16) at (-4, 0.375) {};
		\node [style=none, rotate=90] (17) at (-3.75, 1) {...};
		\node [style=none] (18) at (-0.5, 0.375) {};
		\node [style=none] (19) at (-0.5, 1.625) {};
		\node [style=none] (22) at (-4, -1.625) {};
		\node [style=none] (23) at (-4, -0.375) {};
		\node [style=none] (24) at (0.5, 0.25) {$=$};
		\node [style=none, rotate=90] (25) at (-2.35, 0.05) {...};
		\node [style=none] (26) at (1.5, 1.25) {};
		\node [style=none, rotate=90] (28) at (2, 0) {...};
		\node [style=none] (29) at (5, -1.25) {};
		\node [style=none, rotate=90] (30) at (4.75, 0) {...};
		\node [style=Z phase dot] (32) at (3.25, 0) {$\ \alpha\!+\!\beta\ $};
		\node [style=none] (33) at (5, 1.25) {};
		\node [style=none] (34) at (1.5, -1.25) {};
		\node [style=none] (35) at (0.5, 1) {\spiderrule};
		\node [style=Z phase dot] (36) at (2.75, -4.25) {$\ -\alpha\ $};
		\node [style=none] (37) at (5, -3.125) {};
		\node [style=none] (38) at (0.5, -4.25) {$=$};
		\node [style=none] (39) at (1.5, -4.25) {};
		\node [style=none] (40) at (5, -5.375) {};
		\node [style=X phase dot] (42) at (4.25, -5.375) {$\pi$};
		\node [style=X phase dot] (43) at (-2.75, -4.25) {$\pi$};
		\node [style=none] (44) at (-0.25, -3.125) {};
		\node [style=Z phase dot] (45) at (-1.75, -4.25) {$\alpha$};
		\node [style=none, rotate=90] (46) at (-0.5, -4.25) {...};
		\node [style=none, rotate=90] (48) at (4.25, -4.25) {...};
		\node [style=none] (50) at (-3.5, -4.25) {};
		\node [style=none] (51) at (-0.25, -5.375) {};
		\node [style=X phase dot] (52) at (4.25, -3.125) {$\pi$};
		\node [style=none] (53) at (0.5, -3.5) {\pirule};
		\node [style=X dot] (54) at (12, -3.375) {};
		\node [style=none, rotate=90] (55) at (12.75, -4.25) {...};
		\node [style=Z phase dot] (57) at (8.25, -4.25) {$\alpha$};
		\node [style=none] (58) at (9.5, -5.125) {};
		\node [style=none] (59) at (10.5, -4.25) {$=$};
		\node [style=none] (60) at (13, -5.125) {};
		\node [style=none] (62) at (13, -3.375) {};
		\node [style=none, rotate=90] (63) at (9.25, -4.25) {...};
		\node [style=X dot] (65) at (12, -5.125) {};
		\node [style=none] (66) at (10.5, -3.5) {\copyrule};
		\node [style=none] (67) at (9.5, -3.375) {};
		\node [style=X dot] (68) at (7.25, -4.25) {};
		\node [style=hadamard] (69) at (7.5, 1) {};
		\node [style=none, rotate=90] (72) at (7.5, 0) {...};
		\node [style=none] (73) at (10.75, 0) {$=$};
		\node [style=hadamard] (74) at (7.5, -1) {};
		\node [style=none] (77) at (7, 1) {};
		\node [style=none] (81) at (11.75, -1) {};
		\node [style=none] (82) at (7, -1) {};
		\node [style=none, rotate=90] (83) at (12, 0) {...};
		\node [style=none] (84) at (11.75, 1) {};
		\node [style=none] (86) at (10.75, 0.75) {\hadamardrule};
		\node [style=Z dot] (87) at (16.5, 0.75) {};
		\node [style=none] (88) at (15.75, 0.75) {};
		\node [style=none] (89) at (17.25, 0.75) {};
		\node [style=none] (90) at (19.25, 0.75) {};
		\node [style=none] (91) at (20.25, 0.75) {};
		\node [style=none] (92) at (18.25, 1.5) {\identityrule};
		\node [style=none] (93) at (18.25, 0.75) {$=$};
		\node [style=none] (94) at (18.25, -1.75) {$=$};
		\node [style=none] (95) at (18.25, -1) {\hcancelrule};
		\node [style=none] (96) at (20.25, -1.75) {};
		\node [style=none] (97) at (17.25, -1.75) {};
		\node [style=none] (98) at (15.75, -1.75) {};
		\node [style=none] (99) at (19.25, -1.75) {};
		\node [style=hadamard] (100) at (16.25, -1.75) {};
		\node [style=hadamard] (101) at (16.75, -1.75) {};
		\node [style=none] (102) at (18.25, -3.5) {\bialgrule};
		\node [style=X dot] (103) at (21.25, -5) {};
		\node [style=none] (104) at (22, -3.5) {};
		\node [style=X dot] (105) at (15.5, -4.25) {};
		\node [style=Z dot] (106) at (16.75, -4.25) {};
		\node [style=none] (107) at (19, -5) {};
		\node [style=none] (108) at (14.75, -3.5) {};
		\node [style=none] (109) at (19, -3.5) {};
		\node [style=Z dot] (110) at (19.75, -3.5) {};
		\node [style=none] (111) at (17.5, -3.5) {};
		\node [style=none] (112) at (18.25, -4.25) {$=$};
		\node [style=none] (113) at (14.75, -5) {};
		\node [style=X dot] (114) at (21.25, -3.5) {};
		\node [style=none] (115) at (22, -5) {};
		\node [style=Z dot] (116) at (19.75, -5) {};
		\node [style=none] (117) at (17.5, -5) {};
		\node [style=hadamard] (118) at (9.5, 1) {};
		\node [style=none, rotate=90] (120) at (9.5, 0) {...};
		\node [style=hadamard] (121) at (9.5, -1) {};
		\node [style=Z phase dot] (122) at (8.5, 0) {$\alpha$};
		\node [style=none] (123) at (10, 1) {};
		\node [style=none] (126) at (10, -1) {};
		\node [style=X phase dot] (127) at (13, 0) {$\alpha$};
		\node [style=none] (128) at (14.25, -1) {};
		\node [style=none, rotate=90] (129) at (14, 0) {...};
		\node [style=none] (130) at (14.25, 1) {};
	\end{pgfonlayer}
	\begin{pgfonlayer}{edgelayer}
		\draw [in=-141, out=0, looseness=0.75] (3.center) to (0);
		\draw [in=180, out=-39, looseness=0.75] (0) to (4.center);
		\draw [in=180, out=39, looseness=0.75] (0) to (1.center);
		\draw [in=180, out=0] (2.center) to (0);
		\draw [in=-141, out=0, looseness=0.75] (16.center) to (10);
		\draw [in=180, out=0] (10) to (13.center);
		\draw [in=180, out=39, looseness=0.75] (10) to (11.center);
		\draw [in=141, out=0, looseness=0.75] (12.center) to (10);
		\draw [bend left=45] (10) to (0);
		\draw (11.center) to (19.center);
		\draw (13.center) to (18.center);
		\draw (23.center) to (2.center);
		\draw (22.center) to (3.center);
		\draw [bend left=45] (0) to (10);
		\draw [in=-120, out=0] (34.center) to (32);
		\draw [in=180, out=-60] (32) to (29.center);
		\draw [in=180, out=60] (32) to (33.center);
		\draw [in=120, out=0] (26.center) to (32);
		\draw [in=180, out=-75, looseness=0.75] (45) to (51.center);
		\draw [in=180, out=75, looseness=0.75] (45) to (44.center);
		\draw [in=180, out=0, looseness=0.50] (43) to (45);
		\draw (50.center) to (43);
		\draw [in=-60, out=180, looseness=0.75] (42) to (36);
		\draw [in=0, out=180, looseness=0.75] (36) to (39.center);
		\draw [in=60, out=180, looseness=0.75] (52) to (36);
		\draw (37.center) to (52);
		\draw (40.center) to (42);
		\draw [in=180, out=-75, looseness=0.75] (57) to (58.center);
		\draw [in=180, out=75, looseness=0.75] (57) to (67.center);
		\draw [in=180, out=0, looseness=0.50] (68) to (57);
		\draw (62.center) to (54);
		\draw (60.center) to (65);
		\draw (77.center) to (69);
		\draw (82.center) to (74);
		\draw (88.center) to (89.center);
		\draw (90.center) to (91.center);
		\draw (98.center) to (97.center);
		\draw (99.center) to (96.center);
		\draw (116) to (114);
		\draw (103) to (110);
		\draw (110) to (114);
		\draw (116) to (103);
		\draw (115.center) to (103);
		\draw (107.center) to (116);
		\draw (110) to (109.center);
		\draw (114) to (104.center);
		\draw [bend right] (113.center) to (105);
		\draw [bend right] (105) to (108.center);
		\draw (105) to (106);
		\draw [bend right] (106) to (117.center);
		\draw [bend left] (106) to (111.center);
		\draw [in=-60, out=180, looseness=0.75] (121) to (122);
		\draw [in=75, out=180, looseness=0.75] (118) to (122);
		\draw (123.center) to (118);
		\draw (126.center) to (121);
		\draw [in=-120, out=0, looseness=0.75] (74) to (122);
		\draw [in=105, out=0, looseness=0.75] (69) to (122);
		\draw [in=180, out=-60, looseness=0.75] (127) to (128.center);
		\draw [in=180, out=75, looseness=0.75] (127) to (130.center);
		\draw [in=0, out=-120, looseness=0.75] (127) to (81.center);
		\draw [in=0, out=105, looseness=0.75] (127) to (84.center);
		\draw [in=105, out=-15, looseness=0.75] (10) to (0);
	\end{pgfonlayer}
\end{tikzpicture}

%% file: figures/Zsp-a.tikz
\begin{tikzpicture}
	\begin{pgfonlayer}{nodelayer}
		\node [style=Z phase dot] (0) at (0, 0) {$\alpha$};
		\node [style=none] (1) at (1.25, 1) {};
		\node [style=none] (2) at (-1.25, 1) {};
		\node [style=none] (3) at (-1.25, -1) {};
		\node [style=none] (4) at (1.25, -1) {};
		\node [style=none] (5) at (1.25, 0.5) {};
		\node [style=none] (6) at (-1.25, 0.5) {};
		\node [style=none, rotate=90] (7) at (-1, -0.25) {...};
		\node [style=none, rotate=90] (8) at (1, -0.25) {...};
	\end{pgfonlayer}
	\begin{pgfonlayer}{edgelayer}
		\draw [in=-141, out=0, looseness=0.75] (3.center) to (0);
		\draw [in=180, out=-39, looseness=0.75] (0) to (4.center);
		\draw [in=180, out=22, looseness=0.75] (0) to (5.center);
		\draw [in=180, out=39, looseness=0.75] (0) to (1.center);
		\draw [in=0, out=158, looseness=0.75] (0) to (6.center);
		\draw [in=141, out=0, looseness=0.75] (2.center) to (0);
	\end{pgfonlayer}
\end{tikzpicture}

%% file: figures/Xsp-a.tikz
\begin{tikzpicture}
	\begin{pgfonlayer}{nodelayer}
		\node [style=X phase dot] (0) at (0, 0) {$\alpha$};
		\node [style=none] (1) at (1.25, 1) {};
		\node [style=none] (2) at (-1.25, 1) {};
		\node [style=none] (3) at (-1.25, -1) {};
		\node [style=none] (4) at (1.25, -1) {};
		\node [style=none] (5) at (1.25, 0.5) {};
		\node [style=none] (6) at (-1.25, 0.5) {};
		\node [style=none, rotate=90] (7) at (-1, -0.25) {...};
		\node [style=none, rotate=90] (8) at (1, -0.25) {...};
	\end{pgfonlayer}
	\begin{pgfonlayer}{edgelayer}
		\draw [in=-141, out=0, looseness=0.75] (3.center) to (0);
		\draw [in=180, out=-39, looseness=0.75] (0) to (4.center);
		\draw [in=180, out=22, looseness=0.75] (0) to (5.center);
		\draw [in=180, out=39, looseness=0.75] (0) to (1.center);
		\draw [in=0, out=158, looseness=0.75] (0) to (6.center);
		\draw [in=141, out=0, looseness=0.75] (2.center) to (0);
	\end{pgfonlayer}
\end{tikzpicture}

%% file: figures/ket-+.tikz
\begin{tikzpicture}
	\begin{pgfonlayer}{nodelayer}
		\node [style=Z dot] (0) at (-0.5, 0) {};
		\node [style=none] (1) at (0.5, 0) {};
	\end{pgfonlayer}
	\begin{pgfonlayer}{edgelayer}
		\draw (0) to (1.center);
	\end{pgfonlayer}
\end{tikzpicture}

%% file: figures/ket-0.tikz
\begin{tikzpicture}
	\begin{pgfonlayer}{nodelayer}
		\node [style=X dot] (0) at (-0.5, 0) {};
		\node [style=none] (1) at (0.5, 0) {};
	\end{pgfonlayer}
	\begin{pgfonlayer}{edgelayer}
		\draw (0) to (1.center);
	\end{pgfonlayer}
\end{tikzpicture}

%% file: figures/Z.tikz
\begin{tikzpicture}
	\begin{pgfonlayer}{nodelayer}
		\node [style=Z phase dot] (0) at (0, 0) {$\pi$};
		\node [style=none] (1) at (-1.25, 0) {};
		\node [style=none] (2) at (1.25, 0) {};
	\end{pgfonlayer}
	\begin{pgfonlayer}{edgelayer}
		\draw (1.center) to (0);
		\draw (0) to (2.center);
	\end{pgfonlayer}
\end{tikzpicture}

%% file: figures/X.tikz
\begin{tikzpicture}
	\begin{pgfonlayer}{nodelayer}
		\node [style=X phase dot] (0) at (0, 0) {$\pi$};
		\node [style=none] (1) at (-1.25, 0) {};
		\node [style=none] (2) at (1.25, 0) {};
	\end{pgfonlayer}
	\begin{pgfonlayer}{edgelayer}
		\draw (1.center) to (0);
		\draw (0) to (2.center);
	\end{pgfonlayer}
\end{tikzpicture}

%% file: figures/had-alt.tikz
\begin{tikzpicture}
	\begin{pgfonlayer}{nodelayer}
		\node [style=none] (0) at (-2.75, 0) {};
		\node [style=none] (1) at (-0.75, 0) {};
		\node [style=hadamard] (2) at (-1.75, 0) {};
		\node [style=none] (3) at (0, 0) {$=$};
		\node [style=none] (4) at (0.75, 0) {};
		\node [style=Z phase dot] (5) at (1.5, 0) {$\frac\pi2$};
		\node [style=X phase dot] (6) at (2.5, 0) {$\frac\pi2$};
		\node [style=Z phase dot] (7) at (3.5, 0) {$\frac\pi2$};
		\node [style=none] (8) at (4.25, 0) {};
	\end{pgfonlayer}
	\begin{pgfonlayer}{edgelayer}
		\draw (0.center) to (2);
		\draw (2) to (1.center);
		\draw (4.center) to (8.center);
	\end{pgfonlayer}
\end{tikzpicture}

%% file: figures/cnot.tikz
\begin{tikzpicture}
	\begin{pgfonlayer}{nodelayer}
		\node [style=Z dot] (0) at (0, 0.5) {};
		\node [style=none] (1) at (1.25, 0.5) {};
		\node [style=X dot] (2) at (0, -0.5) {};
		\node [style=none] (3) at (-1.25, 0.5) {};
		\node [style=none] (4) at (1.25, -0.5) {};
		\node [style=none] (5) at (-1.25, -0.5) {};
	\end{pgfonlayer}
	\begin{pgfonlayer}{edgelayer}
		\draw (0) to (2);
		\draw (0) to (3.center);
		\draw (0) to (1.center);
		\draw (5.center) to (2);
		\draw (2) to (4.center);
	\end{pgfonlayer}
\end{tikzpicture}

%% file: figures/Z-a.tikz
\begin{tikzpicture}
	\begin{pgfonlayer}{nodelayer}
		\node [style=Z phase dot] (0) at (0, 0) {$\alpha$};
		\node [style=none] (1) at (-1.25, 0) {};
		\node [style=none] (2) at (1.25, 0) {};
	\end{pgfonlayer}
	\begin{pgfonlayer}{edgelayer}
		\draw (1.center) to (0);
		\draw (0) to (2.center);
	\end{pgfonlayer}
\end{tikzpicture}

%% file: figures/h-alone.tikz
\begin{tikzpicture}
	\begin{pgfonlayer}{nodelayer}
		\node [style=none] (0) at (-0.25, 0) {};
		\node [style=none] (1) at (1.75, 0) {};
		\node [style=hadamard] (2) at (0.75, 0) {};
	\end{pgfonlayer}
	\begin{pgfonlayer}{edgelayer}
		\draw (0.center) to (2);
		\draw (2) to (1.center);
	\end{pgfonlayer}
\end{tikzpicture}

%% file: figures/zx_full.tikz
\begin{tikzpicture}
	\begin{pgfonlayer}{nodelayer}
		\node [style=none] (0) at (-5.75, 1) {};
		\node [style=none] (1) at (-5.75, 0) {};
		\node [style=none] (2) at (-5.75, -1) {};
		\node [style=Z phase dot] (3) at (-4.75, 1) {$\frac{\pi}{2}$};
		\node [style=X dot] (4) at (-3.75, 0) {};
		\node [style=Z dot] (5) at (-3.75, 1) {};
		\node [style=Z dot] (6) at (-2.75, 0) {};
		\node [style=X dot] (7) at (-1.75, -1) {};
		\node [style=Z dot] (8) at (-1.75, 0) {};
		\node [style=none] (9) at (-0.75, 1) {};
		\node [style=none] (10) at (-0.75, 0) {};
		\node [style=none] (11) at (-0.75, -1) {};
		\node [style=none] (12) at (0.75, 1) {};
		\node [style=none] (13) at (0.75, 0) {};
		\node [style=none] (14) at (0.75, -1) {};
		\node [style=Z phase dot] (15) at (1.75, 1) {$\frac{\pi}{2}$};
		\node [style=X dot] (16) at (1.75, 0) {};
		\node [style=Z dot] (17) at (2.75, 0) {};
		\node [style=X dot] (18) at (2.75, -1) {};
		\node [style=none] (19) at (3.75, 1) {};
		\node [style=none] (20) at (3.75, 0) {};
		\node [style=none] (21) at (3.75, -1) {};
		\node [style=none] (22) at (0, 0) {=};
	\end{pgfonlayer}
	\begin{pgfonlayer}{edgelayer}
		\draw (0.center) to (3);
		\draw (1.center) to (4);
		\draw (2.center) to (7);
		\draw (3) to (5);
		\draw (4) to (5);
		\draw [style=hadamard edge] (4) to (6);
		\draw (5) to (9.center);
		\draw (6) to (8);
		\draw (7) to (8);
		\draw (7) to (11.center);
		\draw (8) to (10.center);
		\draw (12.center) to (15);
		\draw (13.center) to (16);
		\draw (14.center) to (18);
		\draw (15) to (16);
		\draw (15) to (19.center);
		\draw [style=hadamard edge] (16) to (17);
		\draw (17) to (18);
		\draw (17) to (20.center);
		\draw (18) to (21.center);
	\end{pgfonlayer}
\end{tikzpicture}

%% file: figures/graph-like-ex.tikz
\begin{tikzpicture}
	\begin{pgfonlayer}{nodelayer}
		\node [style=none] (0) at (-7, 2) {};
		\node [style=none] (1) at (-7, 1) {};
		\node [style=none] (2) at (-7, 0) {};
		\node [style=none] (3) at (-7, -1) {};
		\node [style=none] (4) at (-7, -2) {};
		\node [style=X dot] (5) at (-6, -2) {};
		\node [style=Z dot] (6) at (-6, 2) {};
		\node [style=X dot] (7) at (-5, 2) {};
		\node [style=Z dot] (8) at (-5, -1) {};
		\node [style=Z phase dot] (9) at (-5, -2) {$\frac{\pi}{4}$};
		\node [style=Z phase dot] (10) at (-4, -2) {$\frac{\pi}{4}$};
		\node [style=X dot] (11) at (-4, 0) {};
		\node [style=Z dot] (12) at (-4, 2) {};
		\node [style=Z phase dot] (13) at (-3, 2) {$\frac{\pi}{4}$};
		\node [style=Z dot] (14) at (-4, -1) {};
		\node [style=Z phase dot] (15) at (-3, -1) {$\frac{\pi}{4}$};
		\node [style=X dot] (16) at (-3, 1) {};
		\node [style=Z dot] (17) at (-3, 0) {};
		\node [style=X dot] (18) at (-2, 0) {};
		\node [style=Z dot] (19) at (-2, -2) {};
		\node [style=none] (20) at (-1, 2) {};
		\node [style=none] (21) at (-1, 1) {};
		\node [style=none] (22) at (-1, 0) {};
		\node [style=none] (23) at (-1, -1) {};
		\node [style=none] (24) at (-1, -2) {};
		\node [style=none] (25) at (1, 2) {};
		\node [style=none] (26) at (1, 1) {};
		\node [style=none] (27) at (1, 0) {};
		\node [style=none] (28) at (1, -1) {};
		\node [style=none] (29) at (1, -2) {};
		\node [style=Z dot] (30) at (2, -2) {};
		\node [style=Z dot] (31) at (2, 2) {};
		\node [style=Z dot] (32) at (3, 2) {};
		\node [style=Z dot] (33) at (3, -1) {};
		\node [style=Z phase dot] (34) at (3, -2) {$\frac{\pi}{2}$};
		\node [style=Z dot] (35) at (4, 0) {};
		\node [style=Z phase dot] (36) at (4, 2) {$\frac{\pi}{4}$};
		\node [style=Z phase dot] (37) at (4, -1) {$\frac{\pi}{4}$};
		\node [style=Z dot] (38) at (5, 1) {};
		\node [style=Z dot] (39) at (5, 0) {};
		\node [style=Z dot] (40) at (6, 0) {};
		\node [style=none] (41) at (7, 2) {};
		\node [style=none] (42) at (7, 1) {};
		\node [style=none] (43) at (7, 0) {};
		\node [style=none] (44) at (7, -1) {};
		\node [style=none] (45) at (7, -2) {};
		\node [style=Z dot] (46) at (2.5, 1) {};
		\node [style=hadamard] (48) at (2.5, 0) {};
		\node [style=hadamard] (49) at (1.5, -2) {};
		\node [style=hadamard] (50) at (6, 1) {};
		\node [style=hadamard] (51) at (6.5, 0) {};
		\node [style=none] (52) at (0, 0) {=};
	\end{pgfonlayer}
	\begin{pgfonlayer}{edgelayer}
		\draw (0.center) to (6);
		\draw (1.center) to (16);
		\draw (2.center) to (11);
		\draw (3.center) to (8);
		\draw (4.center) to (5);
		\draw (5) to (6);
		\draw (5) to (9);
		\draw (6) to (7);
		\draw (7) to (8);
		\draw (7) to (12);
		\draw [style=hadamard edge] (8) to (14);
		\draw (9) to (10);
		\draw (10) to (19);
		\draw (11) to (12);
		\draw (11) to (17);
		\draw (12) to (13);
		\draw (13) to (20.center);
		\draw (14) to (15);
		\draw (15) to (23.center);
		\draw (16) to (17);
		\draw (16) to (21.center);
		\draw (17) to (18);
		\draw (18) to (19);
		\draw (18) to (22.center);
		\draw (19) to (24.center);
		\draw (25.center) to (31);
		\draw (26.center) to (46);
		\draw (27.center) to (35);
		\draw (28.center) to (33);
		\draw (29.center) to (30);
		\draw [style=hadamard edge] (30) to (31);
		\draw [style=hadamard edge] (30) to (34);
		\draw [style=hadamard edge] (31) to (32);
		\draw [style=hadamard edge] (32) to (33);
		\draw [style=hadamard edge] (32) to (36);
		\draw [style=hadamard edge] (33) to (37);
		\draw (34) to (45.center);
		\draw [style=hadamard edge] (34) to (40);
		\draw [style=hadamard edge] (35) to (36);
		\draw [style=hadamard edge] (35) to (39);
		\draw (36) to (41.center);
		\draw (37) to (44.center);
		\draw [style=hadamard edge] (38) to (39);
		\draw (38) to (42.center);
		\draw [style=hadamard edge] (39) to (40);
		\draw (40) to (43.center);
		\draw (46) to (38);
	\end{pgfonlayer}
\end{tikzpicture}

%% file: figures/graph1-lab.tikz
\begin{tikzpicture}
	\begin{pgfonlayer}{nodelayer}
		\node [style=vertex] (0) at (-1, 1) {};
		\node [style=vertex] (1) at (-1, -1) {};
		\node [style=vertex] (2) at (1, -1) {};
		\node [style=vertex] (3) at (1, 1) {};
		\node [style=none] (4) at (-1.5, 1) {$a$};
		\node [style=none] (5) at (1.5, 1) {$b$};
		\node [style=none] (6) at (1.5, -1) {$d$};
		\node [style=none] (7) at (-1.5, -1) {$c$};
	\end{pgfonlayer}
	\begin{pgfonlayer}{edgelayer}
		\draw (0) to (3);
		\draw (0) to (2);
		\draw (3) to (2);
		\draw (0) to (1);
	\end{pgfonlayer}
\end{tikzpicture}

%% file: figures/graph1-lab-1.tikz
\begin{tikzpicture}
	\begin{pgfonlayer}{nodelayer}
		\node [style=vertex] (0) at (-1, 1) {};
		\node [style=vertex] (1) at (-1, -1) {};
		\node [style=vertex] (2) at (1, -1) {};
		\node [style=vertex] (3) at (1, 1) {};
		\node [style=none] (4) at (-1.5, 1) {$a$};
		\node [style=none] (5) at (1.5, 1) {$b$};
		\node [style=none] (6) at (1.5, -1) {$d$};
		\node [style=none] (7) at (-1.5, -1) {$c$};
	\end{pgfonlayer}
	\begin{pgfonlayer}{edgelayer}
		\draw (0) to (3);
		\draw (0) to (2);
		\draw (0) to (1);
		\draw (3) to (1);
		\draw (1) to (2);
	\end{pgfonlayer}
\end{tikzpicture}

%% file: figures/local-comp-ex.tikz
\begin{tikzpicture}
	\begin{pgfonlayer}{nodelayer}
		\node [style=Z dot] (0) at (-5.75, 0.5) {};
		\node [style=X phase dot] (1) at (-5.75, 1.5) {-$\frac\pi2$};
		\node [style=Z phase dot] (2) at (-1, 1.5) {$\frac\pi2$};
		\node [style=Z phase dot] (3) at (-2.5, 1.5) {$\frac\pi2$};
		\node [style=Z phase dot] (4) at (-3.5, 1.5) {$\frac\pi2$};
		\node [style=none] (5) at (-1, 2.25) {};
		\node [style=none] (6) at (-2.5, 2.25) {};
		\node [style=none] (7) at (-3.5, 2.25) {};
		\node [style=none] (8) at (-5.75, 2.25) {};
		\node [style=Z dot] (12) at (-2.5, 0.5) {};
		\node [style=Z dot] (13) at (-1, -0.5) {};
		\node [style=Z dot] (14) at (-3.5, -0.5) {};
		\node [style=Z dot] (15) at (-1, -0.5) {};
		\node [style=none] (16) at (-0.5, -1) {};
		\node [style=none] (17) at (-4, -1) {};
		\node [style=none] (18) at (-2.25, -1.5) {$N(u)$};
		\node [style=none] (19) at (-5.75, -0.25) {$u$};
		\node [style=none] (20) at (0.5, 0.5) {$=$};
		\node [style=Z dot] (21) at (2.25, 0.5) {};
		\node [style=Z dot] (22) at (5.5, 0.5) {};
		\node [style=Z dot] (23) at (7, -0.5) {};
		\node [style=Z dot] (24) at (4.5, -0.5) {};
		\node [style=Z dot] (25) at (7, -0.5) {};
		\node [style=none] (30) at (-1.75, 0.75) {...};
		\node [style=none] (31) at (7, 2.25) {};
		\node [style=none] (32) at (5.5, 2.25) {};
		\node [style=none] (33) at (4.5, 2.25) {};
		\node [style=none] (34) at (2.25, 2.25) {};
		\node [style=none] (35) at (6.25, 0.75) {...};
	\end{pgfonlayer}
	\begin{pgfonlayer}{edgelayer}
		\draw (1) to (0);
		\draw (8.center) to (1);
		\draw (2) to (5.center);
		\draw (6.center) to (3);
		\draw (7.center) to (4);
		\draw [style=hadamard edge] (0) to (15);
		\draw (15) to (2);
		\draw (12) to (3);
		\draw (14) to (4);
		\draw [style=hadamard edge] (0) to (14);
		\draw [style=hadamard edge] (0) to (12);
		\draw [style=hadamard edge] (12) to (15);
		\draw [style=brace edge] (16.center) to (17.center);
		\draw [style=hadamard edge] (21) to (25);
		\draw [style=hadamard edge] (21) to (24);
		\draw [style=hadamard edge] (21) to (22);
		\draw (21) to (34.center);
		\draw (24) to (33.center);
		\draw (22) to (32.center);
		\draw (25) to (31.center);
		\draw [style=hadamard edge] (24) to (22);
		\draw [style=hadamard edge] (24) to (25);
	\end{pgfonlayer}
\end{tikzpicture}

%% file: figures/pivot-L.tikz
\begin{tikzpicture}
	\begin{pgfonlayer}{nodelayer}
		\node [style=vertex] (0) at (-2, 1.5) {};
		\node [style=vertex] (1) at (2, 1.5) {};
		\node [style=vertex set] (2) at (0, 0.5) {$~~A~~$};
		\node [style=vertex set] (3) at (-2.5, -0.75) {$~~B~~$};
		\node [style=vertex set] (4) at (2.5, -0.75) {$~~C~~$};
		\node [style=none] (5) at (2.75, 1.5) {$v$};
		\node [style=none] (6) at (-2.75, 1.5) {$u$};
		\node [style=none] (7) at (-0.25, 0.5) {};
		\node [style=none] (8) at (0, 0.25) {};
		\node [style=none] (9) at (0.25, 0.25) {};
		\node [style=none] (10) at (0.5, 0.5) {};
		\node [style=none] (11) at (-2.5, -0.5) {};
		\node [style=none] (12) at (-2.25, -0.75) {};
		\node [style=none] (13) at (-2.25, -0.75) {};
		\node [style=none] (14) at (-2.25, -1) {};
		\node [style=none] (15) at (2.25, -0.75) {};
		\node [style=none] (16) at (2.25, -1) {};
		\node [style=none] (17) at (2.25, -0.75) {};
		\node [style=none] (18) at (2.5, -0.5) {};
	\end{pgfonlayer}
	\begin{pgfonlayer}{edgelayer}
		\draw (0) to (1);
		\draw (0) to (3);
		\draw (2) to (1);
		\draw (0) to (2);
		\draw (1) to (4);
		\draw (7.center) to (11.center);
		\draw (13.center) to (8.center);
		\draw (9.center) to (15.center);
		\draw (10.center) to (18.center);
		\draw (17.center) to (12.center);
		\draw (14.center) to (16.center);
	\end{pgfonlayer}
\end{tikzpicture}

%% file: figures/pivot-R.tikz
\begin{tikzpicture}
	\begin{pgfonlayer}{nodelayer}
		\node [style=vertex] (0) at (-2, 1.5) {};
		\node [style=vertex] (1) at (2, 1.5) {};
		\node [style=vertex set] (2) at (0, 0.5) {$~~A~~$};
		\node [style=vertex set] (3) at (-2.5, -1) {$~~B~~$};
		\node [style=vertex set] (4) at (2.25, -1) {$~~C~~$};
		\node [style=none] (5) at (-3, 1.5) {$v$};
		\node [style=none] (6) at (3, 1.5) {$u$};
		\node [style=none] (7) at (-2.75, -0.75) {};
		\node [style=none] (8) at (-2.25, -1.25) {};
		\node [style=none] (9) at (-0.25, 0.75) {};
		\node [style=none] (10) at (0.25, 0.25) {};
		\node [style=none] (11) at (-0.25, 0.25) {};
		\node [style=none] (12) at (0.25, 0.75) {};
		\node [style=none] (13) at (2.5, -0.75) {};
		\node [style=none] (14) at (2, -1.25) {};
		\node [style=none] (15) at (2.25, -1.25) {};
		\node [style=none] (16) at (2.25, -0.75) {};
		\node [style=none] (17) at (-2.5, -1.25) {};
		\node [style=none] (18) at (-2.5, -0.75) {};
	\end{pgfonlayer}
	\begin{pgfonlayer}{edgelayer}
		\draw (0) to (1);
		\draw (0) to (3);
		\draw (2) to (1);
		\draw (0) to (2);
		\draw (1) to (4);
		\draw (7.center) to (10.center);
		\draw (9.center) to (8.center);
		\draw (18.center) to (15.center);
		\draw (17.center) to (16.center);
		\draw (12.center) to (14.center);
		\draw (11.center) to (13.center);
	\end{pgfonlayer}
\end{tikzpicture}

%% file: figures/pivot-desc.tikz
\begin{tikzpicture}
	\begin{pgfonlayer}{nodelayer}
		\node [style=Z dot] (0) at (-8.75, 1.25) {};
		\node [style=Z dot] (1) at (-3, 1.25) {};
		\node [style=Z dot] (2) at (-6.5, 0.5) {};
		\node [style=Z dot] (3) at (-5.5, -0.75) {};
		\node [style=Z dot] (4) at (-10.25, -1.5) {};
		\node [style=Z dot] (5) at (-9.25, -2.5) {};
		\node [style=Z dot] (6) at (-2.5, -2.5) {};
		\node [style=Z dot] (7) at (-1.5, -1.5) {};
		\node [style=none] (8) at (-8.75, 2.75) {};
		\node [style=none] (9) at (-3, 2.75) {};
		\node [style=none] (10) at (-5.5, 2.75) {};
		\node [style=none] (11) at (-6.5, 2.75) {};
		\node [style=none] (12) at (-10.5, -1) {};
		\node [style=none] (13) at (-9.5, -2) {};
		\node [style=none] (14) at (-2.25, -2) {};
		\node [style=none] (15) at (-1.25, -1) {};
		\node [style=none] (17) at (-1.75, -2.25) {...};
		\node [style=none] (18) at (-10, -2.25) {...};
		\node [style=hadamard] (19) at (-8.75, 2) {};
		\node [style=hadamard] (20) at (-3, 2) {};
		\node [style=Z dot] (21) at (3.25, 1.75) {};
		\node [style=Z dot] (22) at (8.25, 1.75) {};
		\node [style=Z dot] (23) at (5.25, 0.75) {};
		\node [style=Z dot] (24) at (6.25, -0.75) {};
		\node [style=Z dot] (25) at (1.75, -1.25) {};
		\node [style=Z dot] (26) at (2.75, -2.25) {};
		\node [style=Z dot] (27) at (8.75, -2.25) {};
		\node [style=Z dot] (28) at (9.75, -1.25) {};
		\node [style=none] (29) at (3.25, 3) {};
		\node [style=none] (30) at (8.25, 3) {};
		\node [style=none] (31) at (6.25, 3) {};
		\node [style=none] (32) at (5.25, 3) {};
		\node [style=none] (33) at (1.5, -0.75) {};
		\node [style=none] (34) at (2.5, -1.75) {};
		\node [style=none] (35) at (9, -1.75) {};
		\node [style=none] (36) at (10, -0.75) {};
		\node [style=Z phase dot] (40) at (-5.5, 2) {$\pi$};
		\node [style=Z phase dot] (41) at (-6.5, 2) {$\pi$};
		\node [style=none] (42) at (-6, 2.5) {...};
		\node [style=none] (43) at (-6, 0) {...};
		\node [style=none] (44) at (5.75, 0) {...};
		\node [style=none] (45) at (9.75, -2) {...};
		\node [style=none] (46) at (2, -2) {...};
		\node [style=none] (47) at (5.75, 3) {...};
		\node [style=none] (49) at (-9.5, 1.5) {$u$};
		\node [style=none] (50) at (-2.25, 1.5) {$v$};
		\node [style=none] (51) at (-8.5, -3) {};
		\node [style=none] (52) at (-11, -3) {};
		\node [style=none] (53) at (-9.75, -3.5) {$N(u) \backslash (N(v) \cup \{v\})$};
		\node [style=none] (54) at (-0.75, -3) {};
		\node [style=none] (55) at (-3.25, -3) {};
		\node [style=none] (56) at (-2, -3.5) {$N(v) \backslash (N(u) \cup \{u\})$};
		\node [style=none] (57) at (-6.75, 3) {};
		\node [style=none] (58) at (-5.25, 3) {};
		\node [style=none] (59) at (-6, 3.5) {$N(u) \cap N(v)$};
		\node [style=none] (60) at (0, 0) {$=$};
	\end{pgfonlayer}
	\begin{pgfonlayer}{edgelayer}
		\draw (8.center) to (0);
		\draw (9.center) to (1);
		\draw (2) to (11.center);
		\draw (3) to (10.center);
		\draw (6) to (14.center);
		\draw (7) to (15.center);
		\draw (5) to (13.center);
		\draw (4) to (12.center);
		\draw [style=hadamard edge] (0) to (2);
		\draw [style=hadamard edge] (0) to (3);
		\draw [style=hadamard edge] (3) to (1);
		\draw [style=hadamard edge] (2) to (1);
		\draw [style=hadamard edge] (0) to (4);
		\draw [style=hadamard edge] (0) to (5);
		\draw [style=hadamard edge] (1) to (6);
		\draw [style=hadamard edge] (1) to (7);
		\draw [style=hadamard edge] (0) to (1);
		\draw (23) to (32.center);
		\draw (24) to (31.center);
		\draw (27) to (35.center);
		\draw (28) to (36.center);
		\draw (26) to (34.center);
		\draw (25) to (33.center);
		\draw [style=hadamard edge] (21) to (23);
		\draw [style=hadamard edge] (21) to (24);
		\draw [style=hadamard edge] (24) to (22);
		\draw [style=hadamard edge] (23) to (22);
		\draw [style=hadamard edge] (21) to (25);
		\draw [style=hadamard edge] (21) to (26);
		\draw [style=hadamard edge] (22) to (27);
		\draw [style=hadamard edge] (22) to (28);
		\draw [style=hadamard edge] (25) to (24);
		\draw [style=hadamard edge] (26) to (23);
		\draw [style=hadamard edge] (24) to (28);
		\draw [style=hadamard edge] (25) to (27);
		\draw [style=hadamard edge] (26) to (28);
		\draw [in=90, out=-90, looseness=0.50] (29.center) to (22);
		\draw [in=90, out=-90, looseness=0.50] (30.center) to (21);
		\draw [style=hadamard edge] (21) to (22);
		\draw [style=brace edge] (51.center) to (52.center);
		\draw [style=brace edge] (54.center) to (55.center);
		\draw [style=brace edge] (57.center) to (58.center);
		\draw [style=hadamard edge] (5) to (6);
		\draw [style=hadamard edge] (4) to (7);
		\draw [style=hadamard edge] (3) to (5);
		\draw [style=hadamard edge] (2) to (4);
		\draw [style=hadamard edge] (3) to (6);
		\draw [style=hadamard edge] (2) to (7);
		\draw [style=hadamard edge] (27) to (23);
	\end{pgfonlayer}
\end{tikzpicture}

%% file: chapters/methods.tex
\chapter[Methods]{Methods} \label{ch:methods}

The goal of any quantum circuit optimization is to reduce the complexity of the input circuit.
A core assumption of circuit optimization in the ZX-calculus is that ZX-diagrams with fewer spiders typically correspond to simpler circuits.
Earlier, however, we noted that local changes (e.g., connectivity) in a ZX-diagram can drastically affect the complexity of the associated circuit obtained via extraction.
However, no existing optimization procedure over ZX-diagrams addresses this and searches this local space.
The \codeword{full_reduce} method described in Section \ref{sec:zx-circ-opt} is the quintessential example of this lost opportunity.
After the two simplification rules can no longer be applied, the final circuit is that which is extracted from the fully simplified ZX-diagram;
however, there may be an equivalent ZX-diagram lurking nearby whose associated circuit is markedly less complex.

In this thesis, we explore the utility of searching this local space of equivalent ZX-diagrams.
To do so, we first need rewrite rules that modify a ZX-diagram in a manner other than spider removal that may reduce circuit complexity.
Indeed, a rule that introduces several additional spiders alongside changes in connectivity may in turn yield a less complex circuit.
We refer to these non-simplification rewrite rules as \emph{congruences}.
Given a set of congruences, we then need a procedure to search the space of equivalent ZX-diagrams generated by an input ZX-diagram and these rules.
We can then devise strategies for incorporating this local search into existing methods. 

We present the methods of our work in this order.
First, we generalize the original (non-simplification) variants of local complementation and pivoting to ZX-diagrams with arbitrary phases and connectivity.
We then describe two search procedures, simulated annealing (SA) and genetic algorithms (GA).
We also define a measure of circuit complexity and discuss candidate objective functions to guide search. 
Lastly, we discuss how this search can be incorporated into existing optimization pipelines.

\section{Congruences}\label{sec:congruences}

The non-simplification versions of local complementation and pivoting presented in Section \ref{sec:zx-circ-opt} embody the desired properties of congruences.
They change the connectivity of the ZX-diagram without introducing an unwieldy number of additional gates, presenting an opportunity for a potential reduction in the number of 2-qubit gates.
However, Equations \ref{eq:gs-local-comp} and \ref{eq:gs-pivot} only apply to spiders with zero phase and a single wire.

We can apply the rules of the ZX-calculus and Equation \ref{eq:gs-local-comp} to generalize local complementation to arbitrary phases ($\alpha_i, \beta_i \in [0, 2 \pi)$):
\begin{spreadlines}{0.8em}
  \begin{align*}
    \input{gen-lc-single/0.tikz} &\stackrel{(\bm f)}{=} \input{gen-lc-single/1.tikz} \\
    &\stackrel{(\ref{eq:gs-local-comp})}{=} \input{gen-lc-single/2.tikz} \\
    &\stackrel{(\bm i1)}{=} \input{gen-lc-single/3.tikz} \\
    &\stackrel{(\bm f)}{=} \input{gen-lc-single/4.tikz} \\
    &\stackrel{(\bm f)}{=} \input{gen-lc-single/5.tikz} \\[0.8em]
    &\stackrel{(\ref{eq:had-short})}{=} \input{gen-lc-single/6.tikz}\stepcounter{equation}\tag{\theequation}\label{eq:gen-phase-lc}
  \end{align*}
\end{spreadlines}
Equation \ref{eq:gen-phase-lc} can be easily extended to apply for an arbitrary number of wires connected to each spider:
\begin{spreadlines}{0.8em}
  \begin{align*}
    \input{gen-lc-mul/0.tikz} &\stackrel{(\bm f)}{=} \input{gen-lc-mul/1.tikz} \\
    &\stackrel{(\ref{eq:gen-phase-lc})}{=} \input{gen-lc-mul/2.tikz} \\
    &\stackrel{(\bm f)}{=} \input{gen-lc-mul/3.tikz}\stepcounter{equation}\tag{C1}\label{eq:gen-io-lc}
  \end{align*}
\end{spreadlines}
Similarly, we can generalize pivoting to arbitrary phases:
\begin{spreadlines}{0.8em}
  \begin{align*}
    \input{gen-pivot-single/0.tikz} &\stackrel{(\bm f)}{=} \input{gen-pivot-single/1.tikz} \\
    &\stackrel{(\ref{eq:gs-pivot})}{=} \input{gen-pivot-single/2.tikz} \\
    &\stackrel{(\bm f)}{=} \input{gen-pivot-single/3.tikz}\stepcounter{equation}\tag{\theequation}\label{eq:gen-phase-pivot}
  \end{align*}
\end{spreadlines}
Again, we can extend this rewrite rule for arbitrary connectivity:
\begin{spreadlines}{0.8em}
  \begin{align*}
    \input{gen-pivot-mul/0.tikz} &\stackrel{(\bm f)}{=} \input{gen-pivot-mul/1.tikz} \\
    &\stackrel{(\ref{eq:gen-phase-pivot})}{=} \input{gen-pivot-mul/2.tikz} \\
    &\stackrel{(\bm f)}{=} \input{gen-pivot-mul/3.tikz}\stepcounter{equation}\tag{C2}\label{eq:gen-io-pivot}
  \end{align*}
\end{spreadlines}
Equations \ref{eq:gen-io-lc} and \ref{eq:gen-io-pivot} will serve as our primary congruences because they maintain a similar number of spiders but, due to their effects on connectivity, can produce circuits of varying complexities.
Equation \ref{eq:gen-io-lc} can be applied to any spider with a degree of more than 1.
Equation \ref{eq:gen-io-pivot} can be applied to any pair of connected spiders.

\section{Search Procedures}

The space of ZX-diagrams generated by an input ZX-diagram and these congruences is combinatorial and therefore cannot be searched exhaustively.
We can instead formulate this search as an optimization problem where the set of isomorphic ZX-diagrams are the states (reachable by congruence applications) that are evaluated by some quantitative measure of circuit complexity.
Here we describe two search procedures as well as several candidate objective functions that either measure circuit complexity directly or use ZX-diagrammatic properties as a proxy.

\subsection{Simulated Annealing}\label{sec:sa}

Simulated annealing is an optimization technique named after the annealing process in metallurgy in which a molten hot metal is cooled in a slow, controlled fashion in order for it to reach its most stable form~\cite{van1987simulated}.
In SA, this notion of slow cooling is interpreted as a slow decrease in the probability of accepting worse solutions.
In this way, the algorithm initially explores a broad region of the the search space and progressively narrows its scope.

More formally, the SA algorithm maintains a current state $s$ and at each step randomly considers some neighboring state $s^*$.
The algorithm then chooses whether or not to replace $s$ with $s^*$ via a Bernoulli random variable parameterized by the \emph{acceptance probability}, denoted $P(s, s^*, T)$. 
The acceptance probability depends on (1) the \emph{energy} (i.e., objective) \emph{function}, and (2) the \emph{temperature}.
The energy function $E(s)$ assigns a score to each state (lower is better).
If $E(s') < E(s)$, then $s$ is always replaced with $s^*$ (i.e., $P(s, s^*, T) = 1$).
Otherwise, $P(s, s^*, T) = exp(-\frac{E(s^*) - E(s)}{T})$ where the temperature $T$ controls the likelihood of moving to a higher energy state.
Note that $P(s, s^*, T)$ is inversely proportional to $E(s^*) - E(s)$ and directly proportional to $T$.
Informally, this means that $s^*$ is more likely to be accepted if it is closer in energy to $s$ and with a higher temperature.
$T$ is initialized to some positive value and progressively decreases to zero (default: $T = 25$).
At each step, $T$ is updated as $T = T * c$ where $c \in [0, 1)$ is the \emph{cooling} parameter (default: $c = 0.005$). 
In this way, SA converges towards a greedy algorithm and is more likely to accept $s^*$ when $E(s^*) > E(s)$ early in search when $T$ is high.
The algorithm terminates when a maximum number of steps $k_{max}$ is reached (default: $k_{max} = 1000$).
Algorithm \ref{alg:sa} provides an overview of this procedure.

\begin{algorithm}[t]
  \caption{Simulated annealing
    \label{alg:sa}}
  \begin{algorithmic}[1]
    \Function{SimAnneal}{$s_0, T, c, k_{max}$}
      \Let{$s$}{$s_0$}
      \For{$i \gets 0 \textrm{ to } k_{max}$}
        \Let{$s^*$}{randomly sampled neighbor of $s$}
        \If{$E(s^*) < E(s)$ {\bf or} $\text{random}(0, 1) < exp(-\frac{E(s^*) - E(s)}{T})$}
          \Let{$s$}{$s^*$}
        \EndIf
        \Let{$T$}{$T * c$}
      \EndFor
      \State \Return{$s$}
    \EndFunction
  \end{algorithmic}
\end{algorithm}

In this work, the state is a ZX-diagram and neighbors are sampled via the probabilistic application of rewrite rules.
For example, to anneal using the rewrite rules described in Section \ref{sec:congruences}, $s^*$ would be sampled by first choosing one of Equation \ref{eq:gen-io-lc} or Equation \ref{eq:gen-io-pivot} and then choosing a spider or pair of connected spiders to which the rule will be applied.
Note that we use Equations \ref{eq:gen-io-lc} and \ref{eq:gen-io-pivot} by default but in theory any set of rewrite rules can be used.

\subsection{Genetic Algorithms}

Genetic algorithms are a class of optimization procedures inspired by the process of natural selection~\cite{sivanandam2008genetic}.
Rather than maintaining a single state (e.g., as in SA), a \emph{population} of candidate solutions is \emph{evolved}.
This evolution is guided by a \emph{fitness function} and operates via biologically inspired notions of \emph{mutation}, \emph{crossover}, and \emph{selection}.
In this work, we define fitness such that lower fitness is better to remain consistent with the energy function described in Section \ref{sec:sa}.

\begin{figure}
\centering
\input{ga.tikz}
\caption{An overview of genetic algorithms.}
\label{fig:ga}
\end{figure}
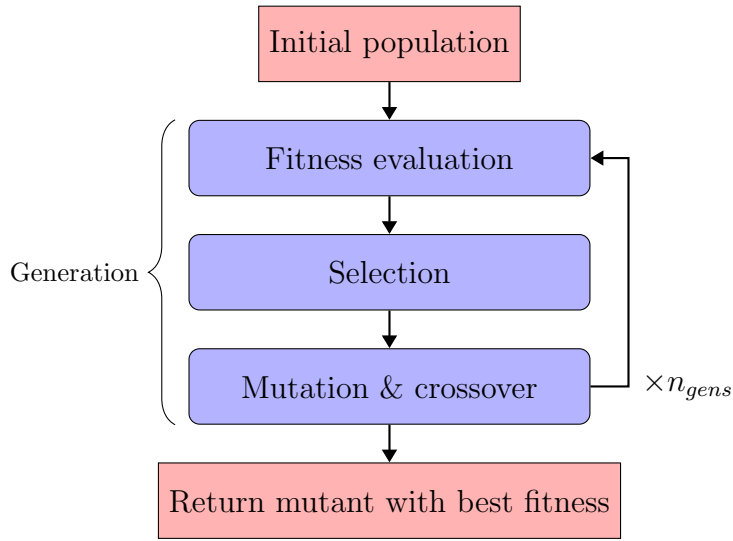

A genetic algorithm operates in the following general fashion.
First, a population of candidate solutions (i.e., \emph{mutants}) is randomly generated.
The population size is typically fixed to some $n_{mutants}$.
This initial population then evolves in a series of \emph{generations}.
Each generation consists of two parts: (1) selection and (2) application of genetic operators (e.g., mutation and crossover).
In the selection step, a score is assigned to each mutant via the fitness function and a new population is chosen given these scores.
For example, a naive selection method would set the population to $n_{mutants}$ copies of the best-scoring mutant.
Given this surviving population, the second step involves applying genetic operators to obtain a revised, diverse population.
The two most common genetic operators are mutation and crossover.
Analogous to mistakes being made during replication of a DNA sequence, a mutation operator involves randomly tweaking a single mutant.
In the case where mutants are represented as bit strings, the canonical example of a mutation operator is a bit-flip.
Alternatively, a crossover operator produces a new mutant given two or more existing mutants;
this is analogous to reproduction and biological crossover.
For example, given two mutants represented as strings, a new mutant can be generated by swapping the two segments defined by a random position.
Evolution proceeds for $n_{gens}$ generations and the algorithm returns the mutant with the highest fitness throughout search.
This process is summarized in Figure \ref{fig:ga}.

In this work, mutants are represented as ZX-diagrams and congruences are used as mutation operators.
As is not uncommon in practical applications of genetic algorithms~\cite{chellapilla1997evolving, spears1991study, de1990genetic}, we do not use any crossover operators though this is an attractive area for future exploration given the graphical nature of ZX-diagrams.
Lastly, we use \emph{tournament selection} to determine the surviving mutants at each generation~\cite{blickle2000tournament}.
In this selection method, $k_{tourn}$ mutants are selected to compete in a ``tournament'' from which a winner is chosen based on fitness.
This is repeated $n_{mutants}$ times with replacement to obtain a surviving population.
We use a simple variant of tournament selection in which $k_{tourn} = 2$ and the winner is simply the mutant with better (i.e., lower) fitness.

\subsection{Objective Functions}\label{sec:obj-funcs}


Both the energy function (used in SA) and the fitness function (used in GA) are instances of an \emph{objective function}.
An objective function measures the ``goodness'' of a particular solution to an optimization problem.
For our purposes, we require an objective function that, given a ZX-diagram, provides a reasonable measure of the complexity of the circuit obtained via extraction.
For consistency, all objective functions are defined in such a way that a low score is better (i.e., a low score indicates low complexity).

The most straightforward objective function involves extracting a circuit and measuring its complexity directly.
A quantitative measure of circuit complexit is also useful for evaluating our optimization techniques.
Given a circuit $C$ in terms of single- and 2-qubit gates, we define its complexity $Comp(C)$ to be a weighted average of its single-qubit and 2-qubit gate counts:
\begin{align*}
  Comp(C) = 10 * (\text{\# 2-qubit gates}) + 1 * (\text{\# single-qubit gates})
\end{align*}
A scaling factor of 10 is chosen because 2-qubit gates are typically an order of magnitude more costly to implement than a single-qubit gate~\cite{campbell2017roads, ballance2016high}.
Note that \codeword{basic_optimization} is always applied to the extracted circuit.


However, extracting a circuit from the ZX-diagram every time we want to measure its complexity is costly.
For example, evolving a population of 50 mutants for 100 generations would require 5000 extractions.
Instead, we could measure some property of the ZX-diagram that is a reasonable proxy for $Comp(C)$ (where $C$ is the circuit obtained via extraction).
The following are candidate properties to serve as such a proxy:
\begin{itemize}
\item
  The number of edges
\item
  The density of the underlying graph~\cite{xu2010web}
\item
  The centrality of the underlying graph~\cite{scott2011sage}
\end{itemize}
The ZX-diagram is assumed to be graph-like.
Note that none of the above objective functions, including $Comp$, are normalized and therefore scores can only be compared between isomorphic ZX-diagrams.

\section{Circuit Optimization Strategy}\label{sec:strategy}

Here we describe how we use these congruences and search procedures for quantum circuit optimization.
At a high-level, our method involves first applying off-the-shelf methods to obtain a simplified ZX-diagram and subsequently searching the space of equivalent ZX-diagrams generated by Equations \ref{eq:gen-io-lc} and \ref{eq:gen-io-pivot}.

\begin{figure}[t]
\centering
\includegraphics[width=15cm]{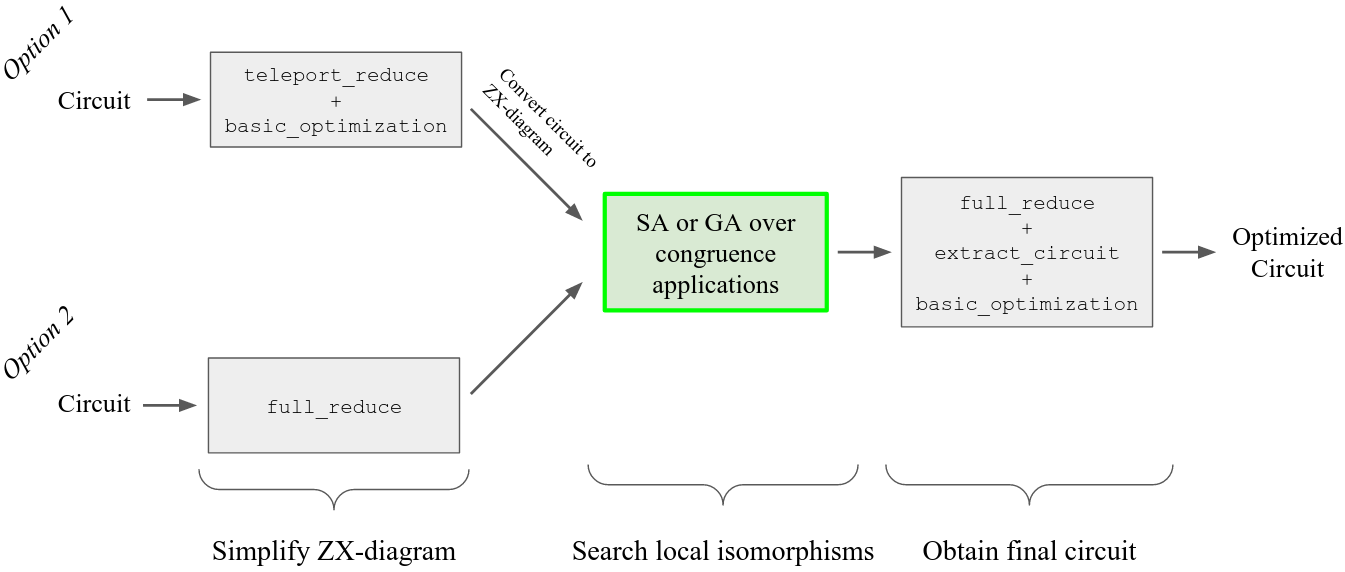}
\caption{
  An overview of our primary optimization strategy.
  Given an input circuit, a simplified ZX-diagram is obtained via one of two off-the-shelf methods.
  Then, we search for an equivalent ZX-diagram with similar graph complexity but lower circuit complexity (highlighted in green).
  A circuit is extracted from the final ZX-diagram and standard circuit-level optimizations are performed.
}
\label{fig:primary}
\end{figure}


Our optimization strategy improves upon the two pipelines described in Section \ref{sec:zx-circ-opt} by searching local variants of the simplified ZX-diagrams.
Given an input circuit, we can obtain a simplified ZX-diagram to seed search in one of two ways.
Firstly, we can apply \codeword{full_reduce};
we do not apply the entire \codeword{full_reduce} + \codeword{extract_circuit} + \codeword{basic_optimization} pipeline and convert the simplified circuit to a ZX-diagram as the complexity cost incurred by extraction will likely outweigh any simplifications achieved at the circuit-level.
Alternatively, since \codeword{teleport_reduce} short-circuits extraction, we can apply \codeword{teleport_reduce} + \codeword{basic_optimization} and convert the simplified circuit to a ZX-diagram.
Given a simplified ZX-diagram, we can then apply SA or GA to search the space of ZX-diagrams generated by Equations \ref{eq:gen-io-lc} and \ref{eq:gen-io-pivot} for an equivalent ZX-diagram whose extracted circuit is less complex.
Note that search proceeds by first randomly selecting one of Equations \ref{eq:gen-io-lc} and \ref{eq:gen-io-pivot} and then randomly selecting a subject (i.e., spider or pair of connected spiders) for the selected congruence.
We denote the probability of selecting Equations \ref{eq:gen-io-lc} and \ref{eq:gen-io-pivot} as $p_{LC}$ and $p_{pivot}$, respectively (default: $p_{LC} = p_{pivot} = 0.5$).
An overview of this strategy is depicted in Figure \ref{fig:primary}.
Importantly, this strategy is unsafe in that the corresponding circuit of the output ZX-diagram can be more complex than that of the input ZX-diagram.
This is particularly a risk when seeding search with a ZX-diagram simplified via \codeword{teleport_reduce} which does not require extraction.

Since Equations \ref{eq:gen-io-lc} and \ref{eq:gen-io-pivot} break the circuit structure of a graph, it is reasonable to apply \codeword{full_reduce} because a circuit has to be extracted regardless.
Therefore, the default objective function scores a ZX-diagram by first applying \codeword{full_reduce}, extracting a circuit $C$ and applying circuit-level optimizations with \codeword{basic_optimization}, and measuring its complexity $Comp(C)$.
We can also apply \codeword{full_reduce} probabilistically via some free parameter $p_{fr}$ throughout search (default: $p_{fr} = 0.1$).
In SA, \codeword{full_reduce} is applied to the current state at each step with probability $p_{fr}$ while in GA \codeword{full_reduce} is applied to each mutant in a given generation with probability $p_{fr}$.

There are several opportunities to refine this strategy.
One example is applying Equations \ref{eq:gen-io-lc} and \ref{eq:gen-io-pivot} with unequal probabilities.
This could be beneficial if one congruence more effectively navigates the search space than the other.
Another possible refinement is non-uniform sampling of spiders (resp. pairs of spiders) to which Equation \ref{eq:gen-io-lc} (resp. Equation \ref{eq:gen-io-pivot}) will be applied.
The following are possible spider metrics to weight sampling for application of Equation \ref{eq:gen-io-lc}:
\begin{itemize}
\item
  Degree
\item
  Centrality measures. For example, the \emph{load} centrality of a node measures the fraction of all shortest paths that pass through that node~\cite{uoh2011universal}
\item
  Degree of neighbors (sum or average)
\end{itemize}
Similarly, the following are possible metrics to weight sampling of pairs of nodes for Equation \ref{eq:gen-io-pivot}:
\begin{itemize}
\item
  Combined degree of the two spiders
\item
  Degree of the union of all neighbors (sum or average)
\item
  Edge centrality measures. For example, the \emph{betweenness} centrality of an edge counts the number of the shortest paths that go through that edge~\cite{brandes2008variants}
\end{itemize}
Given a measure $\mu$ on a set of possible candidates $X$ for congruence application (i.e., spiders or pairs of connected spiders), we compute the probability of choosing some $x \in X$ as
\begin{align*}
  P(x) = \frac{\mu(x)}{\sum_{x' \in X}\mu(x')}
\end{align*}

%% file: figures/gen-lc-single/0.tikz
\begin{tikzpicture}
	\begin{pgfonlayer}{nodelayer}
		\node [style=none] (0) at (2.75, -1.5) {};
		\node [style=none] (1) at (-0.75, -1.5) {};
		\node [style=none] (2) at (1, -2) {$N(u)$};
		\node [style=none] (3) at (-2.5, -0.5) {$u$};
		\node [style=Z phase dot] (4) at (-2.5, 0.25) {$\alpha$};
		\node [style=Z phase dot] (5) at (0.75, 0.25) {$\beta_2$};
		\node [style=Z dot] (6) at (2.25, -0.75) {};
		\node [style=Z phase dot] (7) at (-0.25, -0.75) {$\beta_1$};
		\node [style=Z phase dot] (8) at (2.25, -0.75) {$\beta_3$};
		\node [style=none] (9) at (2.25, 2) {};
		\node [style=none] (10) at (0.75, 2) {};
		\node [style=none] (11) at (-0.25, 2) {};
		\node [style=none] (12) at (-2.5, 2) {};
		\node [style=none] (13) at (1.75, 0.25) {...};
	\end{pgfonlayer}
	\begin{pgfonlayer}{edgelayer}
		\draw [style=brace edge] (0.center) to (1.center);
		\draw [style=hadamard edge] (4) to (8);
		\draw [style=hadamard edge] (4) to (7);
		\draw [style=hadamard edge] (4) to (5);
		\draw (4) to (12.center);
		\draw (7) to (11.center);
		\draw (5) to (10.center);
		\draw (8) to (9.center);
		\draw [style=hadamard edge] (7) to (5);
		\draw [style=hadamard edge] (7) to (8);
	\end{pgfonlayer}
\end{tikzpicture}

%% file: figures/gen-lc-single/1.tikz
\begin{tikzpicture}
	\begin{pgfonlayer}{nodelayer}
		\node [style=Z dot] (0) at (-2.25, -0.25) {};
		\node [style=Z dot] (1) at (1, -0.25) {};
		\node [style=Z dot] (2) at (2.5, -1.25) {};
		\node [style=Z dot] (3) at (0, -1.25) {};
		\node [style=Z dot] (4) at (2.5, -1.25) {};
		\node [style=Z phase dot] (5) at (2.5, 1) {$\beta_3$};
		\node [style=Z phase dot] (6) at (1, 1) {$\beta_2$};
		\node [style=Z phase dot] (7) at (0, 1) {$\beta_1$};
		\node [style=Z phase dot] (8) at (-2.25, 1) {$\alpha$};
		\node [style=none] (9) at (1.75, -0.25) {...};
		\node [style=none] (10) at (-2.25, 2) {};
		\node [style=none] (11) at (0, 2) {};
		\node [style=none] (12) at (1, 2) {};
		\node [style=none] (13) at (2.5, 2) {};
	\end{pgfonlayer}
	\begin{pgfonlayer}{edgelayer}
		\draw [style=hadamard edge] (0) to (4);
		\draw [style=hadamard edge] (0) to (3);
		\draw [style=hadamard edge] (0) to (1);
		\draw (0) to (8);
		\draw (3) to (7);
		\draw (1) to (6);
		\draw (4) to (5);
		\draw [style=hadamard edge] (3) to (1);
		\draw [style=hadamard edge] (3) to (4);
		\draw (8) to (10.center);
		\draw (7) to (11.center);
		\draw (12.center) to (6);
		\draw (13.center) to (5);
	\end{pgfonlayer}
\end{tikzpicture}

%% file: figures/gen-lc-single/2.tikz
\begin{tikzpicture}
	\begin{pgfonlayer}{nodelayer}
		\node [style=none] (0) at (1.75, -1.25) {...};
		\node [style=Z phase dot] (1) at (-2.25, 1) {$\alpha$};
		\node [style=Z phase dot] (2) at (2.75, 1) {$\beta_3$};
		\node [style=Z phase dot] (3) at (-0.25, 1) {$\beta_1$};
		\node [style=Z phase dot] (4) at (0.75, 1) {$\beta_2$};
		\node [style=none] (5) at (-2.25, 2) {};
		\node [style=none] (6) at (-0.25, 2) {};
		\node [style=none] (7) at (0.75, 2) {};
		\node [style=none] (8) at (2.75, 2) {};
		\node [style=Z dot] (9) at (-2.25, -1.5) {};
		\node [style=Z dot] (10) at (-0.25, -2.5) {};
		\node [style=Z dot] (11) at (0.75, -1.5) {};
		\node [style=Z dot] (12) at (2.75, -2.5) {};
		\node [style=X phase dot] (13) at (-2.25, -0.5) {-$\frac\pi2$};
		\node [style=Z phase dot] (14) at (-0.25, -0.5) {$\frac{\pi}{2}$};
		\node [style=Z phase dot] (15) at (0.75, -0.5) {$\frac{\pi}{2}$};
		\node [style=Z phase dot] (16) at (2.75, -0.5) {$\frac{\pi}{2}$};
	\end{pgfonlayer}
	\begin{pgfonlayer}{edgelayer}
		\draw [style=hadamard edge] (9) to (10);
		\draw [style=hadamard edge] (9) to (11);
		\draw [style=hadamard edge] (9) to (12);
		\draw [in=270, out=90] (13) to (1);
		\draw (9) to (13);
		\draw [style=hadamard edge] (11) to (12);
		\draw (14) to (3);
		\draw (10) to (14);
		\draw (11) to (15);
		\draw (15) to (4);
		\draw (16) to (2);
		\draw (12) to (16);
		\draw (8.center) to (2);
		\draw (7.center) to (4);
		\draw (6.center) to (3);
		\draw (5.center) to (1);
	\end{pgfonlayer}
\end{tikzpicture}

%% file: figures/gen-lc-single/3.tikz
\begin{tikzpicture}
	\begin{pgfonlayer}{nodelayer}
		\node [style=none] (0) at (1.75, -1.75) {...};
		\node [style=Z phase dot] (1) at (-2.25, 1.75) {$\alpha$};
		\node [style=Z phase dot] (2) at (2.75, 1.75) {$\beta_3$};
		\node [style=Z phase dot] (3) at (-0.25, 1.75) {$\beta_1$};
		\node [style=Z phase dot] (4) at (0.75, 1.75) {$\beta_2$};
		\node [style=none] (5) at (-2.25, 2.75) {};
		\node [style=none] (6) at (-0.25, 2.75) {};
		\node [style=none] (7) at (0.75, 2.75) {};
		\node [style=none] (8) at (2.75, 2.75) {};
		\node [style=Z dot] (9) at (-2.25, -2) {};
		\node [style=Z dot] (10) at (-0.25, -3) {};
		\node [style=Z dot] (11) at (0.75, -2) {};
		\node [style=Z dot] (12) at (2.75, -3) {};
		\node [style=X phase dot] (13) at (-2.25, -0.25) {-$\frac\pi2$};
		\node [style=Z phase dot] (14) at (-0.25, -0.25) {$\frac{\pi}{2}$};
		\node [style=Z phase dot] (15) at (0.75, -0.25) {$\frac{\pi}{2}$};
		\node [style=Z phase dot] (16) at (2.75, -0.25) {$\frac{\pi}{2}$};
		\node [style=Z dot] (17) at (-2.25, -1) {};
		\node [style=Z dot] (18) at (-2.25, 0.5) {};
	\end{pgfonlayer}
	\begin{pgfonlayer}{edgelayer}
		\draw [style=hadamard edge] (9) to (10);
		\draw [style=hadamard edge] (9) to (11);
		\draw [style=hadamard edge] (9) to (12);
		\draw [style=hadamard edge] (11) to (12);
		\draw (14) to (3);
		\draw (10) to (14);
		\draw (11) to (15);
		\draw (15) to (4);
		\draw (16) to (2);
		\draw (12) to (16);
		\draw (8.center) to (2);
		\draw (7.center) to (4);
		\draw (6.center) to (3);
		\draw (5.center) to (1);
		\draw (1) to (18);
		\draw (18) to (13);
		\draw (13) to (17);
		\draw (17) to (9);
	\end{pgfonlayer}
\end{tikzpicture}

%% file: figures/gen-lc-single/4.tikz
\begin{tikzpicture}
	\begin{pgfonlayer}{nodelayer}
		\node [style=none] (0) at (1.75, -3.25) {...};
		\node [style=Z phase dot] (1) at (-2, 3) {$\alpha$};
		\node [style=Z phase dot] (2) at (2.75, 3) {$\beta_3$};
		\node [style=Z phase dot] (3) at (-0.25, 3) {$\beta_1$};
		\node [style=Z phase dot] (4) at (0.75, 3) {$\beta_2$};
		\node [style=none] (5) at (-2, 4) {};
		\node [style=none] (6) at (-0.25, 4) {};
		\node [style=none] (7) at (0.75, 4) {};
		\node [style=none] (8) at (2.75, 4) {};
		\node [style=Z dot] (9) at (-2, -3.5) {};
		\node [style=Z dot] (10) at (-0.25, -4.5) {};
		\node [style=Z dot] (11) at (0.75, -3.5) {};
		\node [style=Z dot] (12) at (2.75, -4.5) {};
		\node [style=X phase dot] (13) at (-2, -0.5) {-$\frac\pi2$};
		\node [style=Z phase dot] (14) at (-0.25, -0.5) {$\frac{\pi}{2}$};
		\node [style=Z phase dot] (15) at (0.75, -0.5) {$\frac{\pi}{2}$};
		\node [style=Z phase dot] (16) at (2.75, -0.5) {$\frac{\pi}{2}$};
		\node [style=Z phase dot] (17) at (-2, 1.5) {$\frac{\pi}{2}$};
		\node [style=Z phase dot] (18) at (-2, 0.5) {-$\frac{\pi}{2}$};
		\node [style=Z phase dot] (19) at (-2, -1.5) {-$\frac{\pi}{2}$};
		\node [style=Z phase dot] (20) at (-2, -2.5) {$\frac{\pi}{2}$};
	\end{pgfonlayer}
	\begin{pgfonlayer}{edgelayer}
		\draw [style=hadamard edge] (9) to (10);
		\draw [style=hadamard edge] (9) to (11);
		\draw [style=hadamard edge] (9) to (12);
		\draw [style=hadamard edge] (11) to (12);
		\draw (14) to (3);
		\draw (10) to (14);
		\draw (11) to (15);
		\draw (15) to (4);
		\draw (16) to (2);
		\draw (12) to (16);
		\draw (1) to (17);
		\draw (17) to (18);
		\draw (18) to (13);
		\draw (13) to (19);
		\draw (19) to (20);
		\draw (20) to (9);
		\draw (8.center) to (2);
		\draw (7.center) to (4);
		\draw (6.center) to (3);
		\draw (5.center) to (1);
	\end{pgfonlayer}
\end{tikzpicture}

%% file: figures/gen-lc-single/5.tikz
\begin{tikzpicture}
	\begin{pgfonlayer}{nodelayer}
		\node [style=none] (0) at (2.5, -2.25) {...};
		\node [style=Z phase dot] (1) at (-3.25, 2.25) {$\alpha + \frac{\pi}{2}$};
		\node [style=none] (2) at (-3.25, 3.25) {};
		\node [style=none] (3) at (-0.5, 3.25) {};
		\node [style=none] (4) at (1, 3.25) {};
		\node [style=none] (5) at (3.25, 3.25) {};
		\node [style=X phase dot] (6) at (-3.25, 0) {-$\frac\pi2$};
		\node [style=Z phase dot] (7) at (-3.25, 1) {-$\frac{\pi}{2}$};
		\node [style=Z phase dot] (8) at (-3.25, -1) {-$\frac{\pi}{2}$};
		\node [style=Z phase dot] (9) at (-3.25, -2.25) {$\frac{\pi}{2}$};
		\node [style=Z phase dot] (10) at (-0.5, -3.75) {$\beta_1 + \frac{\pi}{2}$};
		\node [style=Z phase dot] (11) at (1, -2) {$\beta_2 + \frac{\pi}{2}$};
		\node [style=Z phase dot] (12) at (3.25, -3.5) {$\beta_3 + \frac{\pi}{2}$};
	\end{pgfonlayer}
	\begin{pgfonlayer}{edgelayer}
		\draw (7) to (6);
		\draw (6) to (8);
		\draw (7) to (1);
		\draw (9) to (8);
		\draw [style=hadamard edge] (9) to (10);
		\draw [style=hadamard edge] (11) to (9);
		\draw [style=hadamard edge] (11) to (12);
		\draw [style=hadamard edge] (9) to (12);
		\draw (2.center) to (1);
		\draw (3.center) to (10);
		\draw (11) to (4.center);
		\draw (12) to (5.center);
	\end{pgfonlayer}
\end{tikzpicture}

%% file: figures/gen-lc-single/6.tikz
\begin{tikzpicture}
	\begin{pgfonlayer}{nodelayer}
		\node [style=none] (0) at (2.75, -0.75) {...};
		\node [style=Z phase dot] (1) at (-3, 0.75) {$\alpha + \frac{\pi}{2}$};
		\node [style=none] (2) at (-3, 2.25) {};
		\node [style=none] (3) at (-0.25, 2.25) {};
		\node [style=none] (4) at (1.25, 2.25) {};
		\node [style=none] (5) at (3.5, 2.25) {};
		\node [style=Z phase dot] (6) at (-3, -0.75) {$\frac{\pi}{2}$};
		\node [style=Z phase dot] (7) at (-0.25, -2.25) {$\beta_1 + \frac{\pi}{2}$};
		\node [style=Z phase dot] (8) at (1.25, -0.5) {$\beta_2 + \frac{\pi}{2}$};
		\node [style=Z phase dot] (9) at (3.5, -2) {$\beta_3 + \frac{\pi}{2}$};
	\end{pgfonlayer}
	\begin{pgfonlayer}{edgelayer}
		\draw [style=hadamard edge] (6) to (7);
		\draw [style=hadamard edge] (8) to (6);
		\draw [style=hadamard edge] (8) to (9);
		\draw [style=hadamard edge] (6) to (9);
		\draw [style=hadamard edge] (6) to (1);
		\draw (9) to (5.center);
		\draw (8) to (4.center);
		\draw (7) to (3.center);
		\draw (1) to (2.center);
	\end{pgfonlayer}
\end{tikzpicture}

%% file: figures/gen-lc-mul/0.tikz
\begin{tikzpicture}
	\begin{pgfonlayer}{nodelayer}
		\node [style=none] (0) at (4, -2.5) {};
		\node [style=none] (1) at (-1.75, -2.5) {};
		\node [style=none] (2) at (1.25, -3) {$N(u)$};
		\node [style=none] (3) at (-3.5, -1) {$u$};
		\node [style=none] (4) at (2.75, 2) {};
		\node [style=none] (5) at (0.25, 2) {};
		\node [style=none] (6) at (-2, 2) {};
		\node [style=none] (7) at (-4.25, 2) {};
		\node [style=none] (8) at (2.25, 0.25) {...};
		\node [style=Z phase dot] (9) at (-3.5, -0.25) {$\alpha$};
		\node [style=Z phase dot] (10) at (3.5, -1) {$\beta_3$};
		\node [style=Z phase dot] (11) at (-1.25, -1.75) {$\beta_1$};
		\node [style=Z phase dot] (12) at (1, 0.25) {$\beta_2$};
		\node [style=none] (13) at (-2.75, 2) {};
		\node [style=none] (14) at (-3.5, 1.25) {...};
		\node [style=none] (15) at (-0.5, 2) {};
		\node [style=none] (16) at (-1.25, 0.75) {...};
		\node [style=none] (17) at (1.75, 2) {};
		\node [style=none] (18) at (1, 1.5) {...};
		\node [style=none] (19) at (4.25, 2) {};
		\node [style=none] (20) at (3.5, 0.75) {...};
	\end{pgfonlayer}
	\begin{pgfonlayer}{edgelayer}
		\draw [style=brace edge] (0.center) to (1.center);
		\draw [bend left] (9) to (7.center);
		\draw [style=hadamard edge] (9) to (10);
		\draw [bend left=15] (10) to (4.center);
		\draw [style=hadamard edge] (11) to (9);
		\draw [style=hadamard edge] (11) to (10);
		\draw [bend left=15] (11) to (6.center);
		\draw [style=hadamard edge] (12) to (11);
		\draw [style=hadamard edge] (12) to (9);
		\draw [bend left] (12) to (5.center);
		\draw [bend right] (9) to (13.center);
		\draw [bend right=15] (11) to (15.center);
		\draw [bend right] (12) to (17.center);
		\draw [bend right=15] (10) to (19.center);
	\end{pgfonlayer}
\end{tikzpicture}

%% file: figures/gen-lc-mul/1.tikz
\begin{tikzpicture}
	\begin{pgfonlayer}{nodelayer}
		\node [style=none] (0) at (3, 2) {};
		\node [style=none] (1) at (0.5, 2) {};
		\node [style=none] (2) at (-1.75, 2) {};
		\node [style=none] (3) at (-4, 2) {};
		\node [style=none] (4) at (2.5, -0.5) {...};
		\node [style=Z dot] (5) at (-3.25, 0.5) {};
		\node [style=Z dot] (6) at (3.75, 0.25) {};
		\node [style=Z dot] (7) at (-1, 0.25) {};
		\node [style=Z dot] (8) at (1.25, 1) {};
		\node [style=none] (9) at (-2.5, 2) {};
		\node [style=none] (10) at (-3.25, 1.5) {...};
		\node [style=none] (11) at (-0.25, 2) {};
		\node [style=none] (12) at (-1, 1.5) {...};
		\node [style=none] (13) at (2, 2) {};
		\node [style=none] (14) at (1.25, 1.75) {...};
		\node [style=none] (15) at (4.5, 2) {};
		\node [style=none] (16) at (3.75, 1.25) {...};
		\node [style=Z phase dot] (17) at (-3.25, -1) {$\alpha$};
		\node [style=Z phase dot] (18) at (-1, -2.5) {$\beta_1$};
		\node [style=Z phase dot] (19) at (1.25, -0.5) {$\beta_2$};
		\node [style=Z phase dot] (20) at (3.75, -1.75) {$\beta_3$};
	\end{pgfonlayer}
	\begin{pgfonlayer}{edgelayer}
		\draw [bend left] (5) to (3.center);
		\draw [bend left=15] (6) to (0.center);
		\draw [bend left=15] (7) to (2.center);
		\draw [bend left] (8) to (1.center);
		\draw [bend right] (5) to (9.center);
		\draw [bend right=15] (7) to (11.center);
		\draw [bend right] (8) to (13.center);
		\draw [bend right=15] (6) to (15.center);
		\draw (17) to (5);
		\draw (18) to (7);
		\draw [style=hadamard edge] (17) to (18);
		\draw (19) to (8);
		\draw [style=hadamard edge] (17) to (19);
		\draw [style=hadamard edge] (18) to (19);
		\draw (20) to (6);
		\draw [style=hadamard edge] (17) to (20);
		\draw [style=hadamard edge] (18) to (20);
	\end{pgfonlayer}
\end{tikzpicture}

%% file: figures/gen-lc-mul/2.tikz
\begin{tikzpicture}
	\begin{pgfonlayer}{nodelayer}
		\node [style=none] (0) at (2.75, -1.5) {...};
		\node [style=Z phase dot] (1) at (-3.25, -0.5) {$\alpha + \frac{\pi}{2}$};
		\node [style=Z phase dot] (2) at (-3.25, -2) {$\frac{\pi}{2}$};
		\node [style=Z phase dot] (3) at (-1, -3.5) {$\beta_1 + \frac{\pi}{2}$};
		\node [style=Z phase dot] (4) at (1.25, -1.5) {$\beta_2 + \frac{\pi}{2}$};
		\node [style=Z phase dot] (5) at (3.75, -2.75) {$\beta_3 + \frac{\pi}{2}$};
		\node [style=none] (6) at (3, 3) {};
		\node [style=none] (7) at (0.5, 3) {};
		\node [style=none] (8) at (-1.75, 3) {};
		\node [style=none] (9) at (-4, 3) {};
		\node [style=Z dot] (10) at (-3.25, 1.5) {};
		\node [style=Z dot] (11) at (3.75, 1.25) {};
		\node [style=Z dot] (12) at (-1, 1.25) {};
		\node [style=Z dot] (13) at (1.25, 2) {};
		\node [style=none] (14) at (-2.5, 3) {};
		\node [style=none] (15) at (-3.25, 2.5) {...};
		\node [style=none] (16) at (-0.25, 3) {};
		\node [style=none] (17) at (-1, 2.5) {...};
		\node [style=none] (18) at (2, 3) {};
		\node [style=none] (19) at (1.25, 2.75) {...};
		\node [style=none] (20) at (4.5, 3) {};
		\node [style=none] (21) at (3.75, 2.25) {...};
	\end{pgfonlayer}
	\begin{pgfonlayer}{edgelayer}
		\draw [style=hadamard edge] (2) to (3);
		\draw [style=hadamard edge] (4) to (2);
		\draw [style=hadamard edge] (4) to (5);
		\draw [style=hadamard edge] (2) to (5);
		\draw [style=hadamard edge] (2) to (1);
		\draw [bend left] (10) to (9.center);
		\draw [bend left=15] (11) to (6.center);
		\draw [bend left=15] (12) to (8.center);
		\draw [bend left] (13) to (7.center);
		\draw [bend right] (10) to (14.center);
		\draw [bend right=15] (12) to (16.center);
		\draw [bend right] (13) to (18.center);
		\draw [bend right=15] (11) to (20.center);
		\draw (1) to (10);
		\draw (3) to (12);
		\draw (4) to (13);
		\draw (5) to (11);
	\end{pgfonlayer}
\end{tikzpicture}

%% file: figures/gen-lc-mul/3.tikz
\begin{tikzpicture}
	\begin{pgfonlayer}{nodelayer}
		\node [style=none] (0) at (3, 2.5) {};
		\node [style=none] (1) at (0.5, 2.5) {};
		\node [style=none] (2) at (-1.75, 2.5) {};
		\node [style=none] (3) at (-4, 2.5) {};
		\node [style=none] (4) at (2.5, 0.25) {...};
		\node [style=Z phase dot] (5) at (-3.25, 0.75) {$\alpha + \frac{\pi}{2}$};
		\node [style=none] (6) at (-2.5, 2.5) {};
		\node [style=none] (7) at (-3.25, 2) {...};
		\node [style=none] (8) at (-0.25, 2.5) {};
		\node [style=none] (9) at (-1, 0.25) {...};
		\node [style=none] (10) at (2, 2.5) {};
		\node [style=none] (11) at (1.25, 1.25) {...};
		\node [style=none] (12) at (4.5, 2.5) {};
		\node [style=none] (13) at (3.75, 0.75) {...};
		\node [style=Z phase dot] (14) at (-3.25, -1.25) {$\frac{\pi}{2}$};
		\node [style=Z phase dot] (15) at (-1, -2.75) {$\beta_1 + \frac{\pi}{2}$};
		\node [style=Z phase dot] (16) at (1.25, -0.75) {$\beta_2 + \frac{\pi}{2}$};
		\node [style=Z phase dot] (17) at (3.75, -2) {$\beta_3 + \frac{\pi}{2}$};
	\end{pgfonlayer}
	\begin{pgfonlayer}{edgelayer}
		\draw [bend left] (5) to (3.center);
		\draw [bend right] (5) to (6.center);
		\draw [style=hadamard edge] (14) to (15);
		\draw [bend left=15, looseness=0.75] (15) to (2.center);
		\draw [bend right=15, looseness=0.75] (15) to (8.center);
		\draw [style=hadamard edge] (16) to (14);
		\draw [bend left=15] (16) to (1.center);
		\draw [bend right=15] (16) to (10.center);
		\draw [style=hadamard edge] (16) to (17);
		\draw [style=hadamard edge] (14) to (17);
		\draw [bend left=15, looseness=0.75] (17) to (0.center);
		\draw [bend right=15, looseness=0.75] (17) to (12.center);
		\draw [style=hadamard edge] (14) to (5);
	\end{pgfonlayer}
\end{tikzpicture}

%% file: figures/gen-pivot-single/0.tikz
\begin{tikzpicture}
	\begin{pgfonlayer}{nodelayer}
		\node [style=Z phase dot] (0) at (-2.5, 1.75) {$\alpha_1$};
		\node [style=Z phase dot] (1) at (2.5, 1.75) {$\alpha_2$};
		\node [style=Z phase dot] (2) at (-0.5, 0.75) {$\beta_1$};
		\node [style=Z phase dot] (3) at (0.5, -0.75) {$\beta_2$};
		\node [style=Z phase dot] (4) at (-4, -1.25) {$\beta_3$};
		\node [style=Z phase dot] (5) at (-3, -2.25) {$\beta_4$};
		\node [style=Z phase dot] (6) at (3, -2.25) {$\beta_5$};
		\node [style=Z phase dot] (7) at (4, -1.25) {$\beta_6$};
		\node [style=none] (8) at (-2.5, 3) {};
		\node [style=none] (9) at (2.5, 3) {};
		\node [style=none] (10) at (0.5, 3) {};
		\node [style=none] (11) at (-0.5, 3) {};
		\node [style=none] (12) at (-4.25, -0.75) {};
		\node [style=none] (13) at (-3.25, -1.75) {};
		\node [style=none] (14) at (3.25, -1.75) {};
		\node [style=none] (15) at (4.25, -0.75) {};
		\node [style=none] (16) at (0, 0) {...};
		\node [style=none] (17) at (4, -2) {...};
		\node [style=none] (18) at (-3.75, -2) {...};
		\node [style=none] (19) at (0, 2.75) {...};
		\node [style=none] (20) at (-3.25, 2.25) {$u$};
		\node [style=none] (21) at (3.25, 2.25) {$v$};
		\node [style=none] (22) at (-2.25, -3) {};
		\node [style=none] (23) at (-4.75, -3) {};
		\node [style=none] (24) at (-3.5, -3.5) {${\scriptstyle N(u) \backslash (N(v) \cup \{v\})}$};
		\node [style=none] (25) at (4.75, -3) {};
		\node [style=none] (26) at (2.25, -3) {};
		\node [style=none] (27) at (3.5, -3.5) {${\scriptstyle N(v) \backslash (N(u) \cup \{u\})}$};
		\node [style=none] (28) at (-0.75, 3.25) {};
		\node [style=none] (29) at (0.75, 3.25) {};
		\node [style=none] (30) at (0, 3.75) {${\scriptstyle N(u) \cap N(v)}$};
	\end{pgfonlayer}
	\begin{pgfonlayer}{edgelayer}
		\draw (2) to (11.center);
		\draw (3) to (10.center);
		\draw (6) to (14.center);
		\draw (7) to (15.center);
		\draw (5) to (13.center);
		\draw (4) to (12.center);
		\draw [style=hadamard edge] (0) to (2);
		\draw [style=hadamard edge] (0) to (3);
		\draw [style=hadamard edge] (3) to (1);
		\draw [style=hadamard edge] (2) to (1);
		\draw [style=hadamard edge] (0) to (4);
		\draw [style=hadamard edge] (0) to (5);
		\draw [style=hadamard edge] (1) to (6);
		\draw [style=hadamard edge] (1) to (7);
		\draw [style=hadamard edge] (4) to (3);
		\draw [style=hadamard edge] (5) to (2);
		\draw [style=hadamard edge] (3) to (7);
		\draw [style=hadamard edge] (4) to (6);
		\draw [style=hadamard edge] (5) to (7);
		\draw [in=90, out=-90, looseness=0.50] (8.center) to (1);
		\draw [in=90, out=-90, looseness=0.50] (9.center) to (0);
		\draw [style=hadamard edge] (0) to (1);
		\draw [style=brace edge] (22.center) to (23.center);
		\draw [style=brace edge] (25.center) to (26.center);
		\draw [style=brace edge] (28.center) to (29.center);
		\draw [style=hadamard edge] (6) to (2);
	\end{pgfonlayer}
\end{tikzpicture}

%% file: figures/gen-pivot-single/1.tikz
\begin{tikzpicture}
	\begin{pgfonlayer}{nodelayer}
		\node [style=Z dot] (0) at (-2.5, 2) {};
		\node [style=Z dot] (1) at (2.5, 2) {};
		\node [style=Z dot] (2) at (-0.5, 1) {};
		\node [style=Z dot] (3) at (0.5, -0.75) {};
		\node [style=Z dot] (4) at (-4, -1.5) {};
		\node [style=Z dot] (5) at (-3.75, -3.75) {};
		\node [style=Z dot] (6) at (3.75, -3.75) {};
		\node [style=Z dot] (7) at (4, -1.5) {};
		\node [style=Z phase dot] (8) at (-2.5, 3.25) {$\alpha_2$};
		\node [style=Z phase dot] (9) at (2.5, 3.25) {$\alpha_1$};
		\node [style=Z phase dot] (10) at (0.5, 3.25) {$\beta_2$};
		\node [style=Z phase dot] (11) at (-0.5, 3.25) {$\beta_1$};
		\node [style=Z phase dot] (12) at (-4.5, -0.75) {$\beta_3$};
		\node [style=Z phase dot] (13) at (-4.25, -3) {$\beta_4$};
		\node [style=Z phase dot] (14) at (4.25, -3) {$\beta_5$};
		\node [style=Z phase dot] (15) at (4.5, -0.75) {$\beta_6$};
		\node [style=none] (16) at (-0.2, 0.225) {...};
		\node [style=none] (17) at (4, -2.25) {...};
		\node [style=none] (18) at (-4, -2.25) {...};
		\node [style=none] (19) at (0, 4) {...};
		\node [style=none] (20) at (2.5, 4.5) {};
		\node [style=none] (21) at (-2.5, 4.5) {};
		\node [style=none] (22) at (-0.5, 4.5) {};
		\node [style=none] (23) at (0.5, 4.5) {};
		\node [style=none] (24) at (5, 0) {};
		\node [style=none] (25) at (4.75, -2.25) {};
		\node [style=none] (26) at (-4.75, -2.25) {};
		\node [style=none] (27) at (-5, 0) {};
	\end{pgfonlayer}
	\begin{pgfonlayer}{edgelayer}
		\draw (2) to (11);
		\draw (3) to (10);
		\draw (6) to (14);
		\draw (7) to (15);
		\draw (5) to (13);
		\draw (4) to (12);
		\draw [style=hadamard edge] (0) to (2);
		\draw [style=hadamard edge] (0) to (3);
		\draw [style=hadamard edge] (3) to (1);
		\draw [style=hadamard edge] (2) to (1);
		\draw [style=hadamard edge] (0) to (4);
		\draw [style=hadamard edge] (0) to (5);
		\draw [style=hadamard edge] (1) to (6);
		\draw [style=hadamard edge] (1) to (7);
		\draw [style=hadamard edge] (4) to (3);
		\draw [style=hadamard edge] (5) to (2);
		\draw [style=hadamard edge] (3) to (7);
		\draw [style=hadamard edge] (4) to (6);
		\draw [style=hadamard edge] (5) to (7);
		\draw [in=90, out=-90, looseness=0.50] (8) to (1);
		\draw [in=90, out=-90, looseness=0.50] (9) to (0);
		\draw [style=hadamard edge] (0) to (1);
		\draw [style=hadamard edge] (6) to (2);
		\draw (9) to (20.center);
		\draw (8) to (21.center);
		\draw (11) to (22.center);
		\draw (10) to (23.center);
		\draw (15) to (24.center);
		\draw (14) to (25.center);
		\draw (13) to (26.center);
		\draw (12) to (27.center);
	\end{pgfonlayer}
\end{tikzpicture}

%% file: figures/gen-pivot-single/2.tikz
\begin{tikzpicture}
	\begin{pgfonlayer}{nodelayer}
		\node [style=Z dot] (0) at (-2.5, 1.75) {};
		\node [style=Z dot] (1) at (2.5, 1.75) {};
		\node [style=Z dot] (2) at (-0.5, 0.75) {};
		\node [style=Z dot] (3) at (0.5, -1) {};
		\node [style=Z dot] (4) at (-4, -1.75) {};
		\node [style=Z dot] (5) at (-3.75, -4) {};
		\node [style=Z dot] (6) at (3.75, -4) {};
		\node [style=Z dot] (7) at (4, -1.75) {};
		\node [style=Z phase dot] (8) at (-2.5, 4.25) {$\alpha_2$};
		\node [style=Z phase dot] (9) at (2.5, 4.25) {$\alpha_1$};
		\node [style=Z phase dot] (10) at (0.5, 4.25) {$\beta_2$};
		\node [style=Z phase dot] (11) at (-0.5, 4.25) {$\beta_1$};
		\node [style=Z phase dot] (12) at (-4.5, -1) {$\beta_3$};
		\node [style=Z phase dot] (13) at (-4.25, -3.25) {$\beta_4$};
		\node [style=Z phase dot] (14) at (4.25, -3.25) {$\beta_5$};
		\node [style=Z phase dot] (15) at (4.5, -1) {$\beta_6$};
		\node [style=none] (16) at (-0.075, 0.025) {...};
		\node [style=none] (17) at (4, -2.5) {...};
		\node [style=none] (18) at (-4, -2.5) {...};
		\node [style=none] (19) at (0, 5) {...};
		\node [style=none] (20) at (2.5, 5.5) {};
		\node [style=none] (21) at (-2.5, 5.5) {};
		\node [style=none] (22) at (-0.5, 5.5) {};
		\node [style=none] (23) at (0.5, 5.5) {};
		\node [style=none] (24) at (5, -0.25) {};
		\node [style=none] (25) at (4.75, -2.5) {};
		\node [style=none] (26) at (-4.75, -2.5) {};
		\node [style=none] (27) at (-5, -0.25) {};
		\node [style=hadamard] (28) at (-2.5, 3) {};
		\node [style=hadamard] (29) at (2.5, 3) {};
		\node [style=Z phase dot] (30) at (-0.5, 3) {$\pi$};
		\node [style=Z phase dot] (31) at (0.5, 3) {$\pi$};
	\end{pgfonlayer}
	\begin{pgfonlayer}{edgelayer}
		\draw (6) to (14);
		\draw (7) to (15);
		\draw (5) to (13);
		\draw (4) to (12);
		\draw [style=hadamard edge] (0) to (2);
		\draw [style=hadamard edge] (0) to (3);
		\draw [style=hadamard edge] (3) to (1);
		\draw [style=hadamard edge] (2) to (1);
		\draw [style=hadamard edge] (0) to (4);
		\draw [style=hadamard edge] (0) to (5);
		\draw [style=hadamard edge] (1) to (6);
		\draw [style=hadamard edge] (1) to (7);
		\draw [style=hadamard edge] (0) to (1);
		\draw (9) to (20.center);
		\draw (8) to (21.center);
		\draw (11) to (22.center);
		\draw (10) to (23.center);
		\draw (15) to (24.center);
		\draw (14) to (25.center);
		\draw (13) to (26.center);
		\draw (12) to (27.center);
		\draw (8) to (28);
		\draw (28) to (0);
		\draw (1) to (29);
		\draw (29) to (9);
		\draw (30) to (11);
		\draw (2) to (30);
		\draw (3) to (31);
		\draw (31) to (10);
		\draw [style=hadamard edge] (4) to (2);
		\draw [style=hadamard edge] (5) to (3);
		\draw [style=hadamard edge] (4) to (7);
		\draw [style=hadamard edge] (5) to (6);
		\draw [style=hadamard edge] (2) to (7);
		\draw [style=hadamard edge] (3) to (6);
	\end{pgfonlayer}
\end{tikzpicture}

%% file: figures/gen-pivot-single/3.tikz
\begin{tikzpicture}
	\begin{pgfonlayer}{nodelayer}
		\node [style=Z dot] (0) at (-2.5, 2.25) {};
		\node [style=Z dot] (1) at (2.5, 2.25) {};
		\node [style=Z phase dot] (2) at (-0.5, 1.25) {$\beta_1 + \pi$};
		\node [style=Z phase dot] (3) at (0.5, -1) {$\beta_2 + \pi$};
		\node [style=Z phase dot] (4) at (-4, -2) {$\beta_3$};
		\node [style=Z phase dot] (5) at (-3, -3) {$\beta_4$};
		\node [style=Z phase dot] (6) at (4, -2) {$\beta_6$};
		\node [style=Z phase dot] (7) at (-2.5, 3.5) {$\alpha_2$};
		\node [style=Z phase dot] (8) at (2.5, 3.5) {$\alpha_1$};
		\node [style=none] (9) at (0, 0.25) {...};
		\node [style=none] (10) at (3.75, -2.75) {...};
		\node [style=none] (11) at (-3.75, -2.75) {...};
		\node [style=none] (12) at (0, 3.5) {...};
		\node [style=none] (13) at (2.5, 4.75) {};
		\node [style=none] (14) at (-2.5, 4.75) {};
		\node [style=none] (15) at (-0.5, 4.75) {};
		\node [style=none] (16) at (0.5, 4.75) {};
		\node [style=none] (17) at (4.25, -1.5) {};
		\node [style=none] (18) at (3.25, -2.5) {};
		\node [style=none] (19) at (-3.25, -2.5) {};
		\node [style=none] (20) at (-4.25, -1.5) {};
		\node [style=Z phase dot] (21) at (3, -3) {$\beta_5$};
	\end{pgfonlayer}
	\begin{pgfonlayer}{edgelayer}
		\draw [style=hadamard edge] (0) to (2);
		\draw [style=hadamard edge] (0) to (3);
		\draw [style=hadamard edge] (3) to (1);
		\draw [style=hadamard edge] (2) to (1);
		\draw [style=hadamard edge] (0) to (4);
		\draw [style=hadamard edge] (0) to (5);
		\draw [style=hadamard edge] (1) to (6);
		\draw [style=hadamard edge] (0) to (1);
		\draw (8) to (13.center);
		\draw (7) to (14.center);
		\draw [style=hadamard edge] (4) to (2);
		\draw [style=hadamard edge] (5) to (3);
		\draw [style=hadamard edge] (4) to (6);
		\draw [style=hadamard edge] (2) to (6);
		\draw [style=hadamard edge] (1) to (21);
		\draw [style=hadamard edge] (5) to (21);
		\draw [style=hadamard edge] (3) to (21);
		\draw (21) to (18.center);
		\draw (6) to (17.center);
		\draw (5) to (19.center);
		\draw (4) to (20.center);
		\draw (3) to (16.center);
		\draw (2) to (15.center);
		\draw [style=hadamard edge, in=270, out=90] (1) to (8);
		\draw [style=hadamard edge] (0) to (7);
	\end{pgfonlayer}
\end{tikzpicture}

%% file: figures/gen-pivot-mul/0.tikz
\begin{tikzpicture}
	\begin{pgfonlayer}{nodelayer}
		\node [style=Z phase dot] (0) at (-3.5, 1.75) {$\alpha_1$};
		\node [style=Z phase dot] (1) at (3.5, 1.75) {$\alpha_2$};
		\node [style=Z phase dot] (2) at (-1.25, -0.25) {$\beta_1$};
		\node [style=Z phase dot] (3) at (1.25, -1) {$\beta_2$};
		\node [style=Z phase dot] (4) at (-4.75, -1.5) {$\beta_3$};
		\node [style=Z phase dot] (5) at (-2.5, -3.75) {$\beta_4$};
		\node [style=Z phase dot] (6) at (2.5, -3.75) {$\beta_5$};
		\node [style=Z phase dot] (7) at (4.75, -1.5) {$\beta_6$};
		\node [style=none] (8) at (-2.5, 4) {};
		\node [style=none] (9) at (2.5, 4) {};
		\node [style=none] (10) at (1.75, 4) {};
		\node [style=none] (11) at (-1.75, 4) {};
		\node [style=none] (12) at (-5, -0.25) {};
		\node [style=none] (13) at (-3.5, -3) {};
		\node [style=none] (14) at (3.25, -3) {};
		\node [style=none] (15) at (5, -0.25) {};
		\node [style=none] (16) at (-0.175, -0.675) {...};
		\node [style=none] (17) at (3.5, -2.25) {...};
		\node [style=none] (18) at (-3.5, -2.25) {...};
		\node [style=none] (19) at (0, 2.25) {...};
		\node [style=none] (20) at (-3.5, 4) {};
		\node [style=none] (21) at (3.5, 4) {};
		\node [style=none] (22) at (2.925, 3.75) {...};
		\node [style=none] (23) at (-2.925, 3.75) {...};
		\node [style=none] (24) at (-0.75, 4) {};
		\node [style=none] (25) at (-1.25, 3.5) {...};
		\node [style=none] (26) at (0.75, 4) {};
		\node [style=none] (27) at (1.25, 3.5) {...};
		\node [style=none] (28) at (6, -0.25) {};
		\node [style=none] (29) at (-6, -0.25) {};
		\node [style=none] (30) at (4.5, -3) {};
		\node [style=none] (31) at (-4.75, -3) {};
		\node [style=none] (32) at (5.35, -0.5) {...};
		\node [style=none] (33) at (-5.3, -0.5) {...};
		\node [style=none] (34) at (3.85, -3.025) {...};
		\node [style=none] (35) at (-4.1, -3.05) {...};
		\node [style=none] (36) at (-1.75, 4.75) {};
		\node [style=none] (37) at (1.75, 4.75) {};
		\node [style=none] (38) at (0, 5.25) {${\scriptstyle N(u) \cap N(v)}$};
		\node [style=none] (39) at (-2, -4.5) {};
		\node [style=none] (40) at (-5.25, -4.5) {};
		\node [style=none] (41) at (-3.75, -5) {${\scriptstyle N(u) \backslash (N(v) \cup \{v\})}$};
		\node [style=none] (42) at (5.25, -4.5) {};
		\node [style=none] (43) at (2, -4.5) {};
		\node [style=none] (44) at (3.75, -5) {${\scriptstyle N(v) \backslash (N(u) \cup \{u\})}$};
		\node [style=none] (45) at (-4.25, 2.25) {$u$};
		\node [style=none] (46) at (4.25, 2.25) {$v$};
	\end{pgfonlayer}
	\begin{pgfonlayer}{edgelayer}
		\draw [bend left=15, looseness=0.50] (2) to (11.center);
		\draw [bend right=15, looseness=0.50] (3) to (10.center);
		\draw [bend left=15] (6) to (14.center);
		\draw [bend left=15, looseness=0.75] (7) to (15.center);
		\draw [bend right=15] (5) to (13.center);
		\draw [bend right=15, looseness=0.75] (4) to (12.center);
		\draw [style=hadamard edge] (0) to (2);
		\draw [style=hadamard edge] (0) to (3);
		\draw [style=hadamard edge] (3) to (1);
		\draw [style=hadamard edge] (2) to (1);
		\draw [style=hadamard edge] (0) to (4);
		\draw [style=hadamard edge] (0) to (5);
		\draw [style=hadamard edge] (1) to (6);
		\draw [style=hadamard edge] (1) to (7);
		\draw [style=hadamard edge] (4) to (3);
		\draw [style=hadamard edge] (5) to (2);
		\draw [style=hadamard edge] (3) to (7);
		\draw [style=hadamard edge] (4) to (6);
		\draw [style=hadamard edge] (5) to (7);
		\draw [in=90, out=-90, looseness=0.50] (8.center) to (1);
		\draw [in=90, out=-90, looseness=0.50] (9.center) to (0);
		\draw [style=hadamard edge] (0) to (1);
		\draw [style=hadamard edge] (6) to (2);
		\draw [in=120, out=-90, looseness=0.50] (20.center) to (1);
		\draw [in=-90, out=60, looseness=0.50] (0) to (21.center);
		\draw [bend right=15, looseness=0.50] (2) to (24.center);
		\draw [bend right=15, looseness=0.50] (26.center) to (3);
		\draw [bend right, looseness=0.75] (7) to (28.center);
		\draw [bend left, looseness=0.75] (4) to (29.center);
		\draw [bend left=15] (5) to (31.center);
		\draw [bend right=15] (6) to (30.center);
		\draw [style=brace edge] (36.center) to (37.center);
		\draw [style=brace edge] (39.center) to (40.center);
		\draw [style=brace edge] (42.center) to (43.center);
	\end{pgfonlayer}
\end{tikzpicture}

%% file: figures/gen-pivot-mul/1.tikz
\begin{tikzpicture}
	\begin{pgfonlayer}{nodelayer}
		\node [style=Z dot] (0) at (3.5, 3) {};
		\node [style=Z dot] (1) at (-3.5, 3) {};
		\node [style=Z dot] (2) at (-1.25, 2.75) {};
		\node [style=Z dot] (3) at (1.25, 2.75) {};
		\node [style=Z dot] (4) at (-5.5, -1) {};
		\node [style=Z dot] (5) at (-3.5, -3.75) {};
		\node [style=Z dot] (6) at (3.5, -3.75) {};
		\node [style=Z dot] (7) at (5.5, -1) {};
		\node [style=none] (8) at (-3, 4.25) {};
		\node [style=none] (9) at (3, 4.25) {};
		\node [style=none] (10) at (1.75, 4.25) {};
		\node [style=none] (11) at (-1.75, 4.25) {};
		\node [style=none] (12) at (-5.75, 0.25) {};
		\node [style=none] (13) at (-4.25, -3) {};
		\node [style=none] (14) at (4.25, -3) {};
		\node [style=none] (15) at (5.75, 0.25) {};
		\node [style=none] (16) at (-0.175, -1.175) {...};
		\node [style=none] (17) at (3.5, -2.75) {...};
		\node [style=none] (18) at (-3.5, -2.75) {...};
		\node [style=none] (19) at (0, 2.75) {...};
		\node [style=none] (20) at (-4, 4.25) {};
		\node [style=none] (21) at (4, 4.25) {};
		\node [style=none] (22) at (3.5, 4) {...};
		\node [style=none] (23) at (-3.475, 4) {...};
		\node [style=none] (24) at (-0.75, 4.25) {};
		\node [style=none] (25) at (-1.25, 4) {...};
		\node [style=none] (26) at (0.75, 4.25) {};
		\node [style=none] (27) at (1.25, 4) {...};
		\node [style=none] (28) at (6.75, 0.25) {};
		\node [style=none] (29) at (-6.75, 0.25) {};
		\node [style=none] (30) at (5.5, -3) {};
		\node [style=none] (31) at (-5.5, -3) {};
		\node [style=none] (32) at (6.15, -0.025) {...};
		\node [style=none] (33) at (-6.1, 0) {...};
		\node [style=none] (34) at (4.75, -3.025) {...};
		\node [style=none] (35) at (-4.75, -3.05) {...};
		\node [style=Z phase dot] (36) at (3.5, 1.25) {$\alpha_2$};
		\node [style=Z phase dot] (37) at (-3.5, 1.25) {$\alpha_1$};
		\node [style=Z phase dot] (38) at (4.75, -2) {$\beta_6$};
		\node [style=Z phase dot] (39) at (-1.25, -0.75) {$\beta_1$};
		\node [style=Z phase dot] (40) at (1.25, -1.5) {$\beta_2$};
		\node [style=Z phase dot] (41) at (-4.75, -2) {$\beta_3$};
		\node [style=Z phase dot] (42) at (2.5, -4.25) {$\beta_5$};
		\node [style=Z phase dot] (43) at (-2.5, -4.25) {$\beta_4$};
	\end{pgfonlayer}
	\begin{pgfonlayer}{edgelayer}
		\draw [bend left=15, looseness=0.75] (2) to (11.center);
		\draw [bend right=15, looseness=0.75] (3) to (10.center);
		\draw [bend left=15] (6) to (14.center);
		\draw [bend left=15, looseness=0.75] (7) to (15.center);
		\draw [bend right=15] (5) to (13.center);
		\draw [bend right=15, looseness=0.75] (4) to (12.center);
		\draw [bend right=15, looseness=0.75] (2) to (24.center);
		\draw [bend right=15, looseness=0.75] (26.center) to (3);
		\draw [bend right, looseness=0.75] (7) to (28.center);
		\draw [bend left, looseness=0.75] (4) to (29.center);
		\draw [bend left=15] (5) to (31.center);
		\draw [bend right=15] (6) to (30.center);
		\draw [in=-90, out=90, looseness=0.50] (36) to (1);
		\draw [style=hadamard edge] (37) to (36);
		\draw [in=-90, out=90, looseness=0.50] (37) to (0);
		\draw [style=hadamard edge] (38) to (36);
		\draw (38) to (7);
		\draw [style=hadamard edge] (39) to (37);
		\draw [style=hadamard edge] (39) to (36);
		\draw (39) to (2);
		\draw [style=hadamard edge] (40) to (36);
		\draw [style=hadamard edge] (40) to (38);
		\draw [style=hadamard edge] (40) to (37);
		\draw (40) to (3);
		\draw [style=hadamard edge] (41) to (37);
		\draw [style=hadamard edge] (41) to (40);
		\draw (41) to (4);
		\draw [bend right] (0) to (21.center);
		\draw [bend left] (0) to (9.center);
		\draw [bend right] (1) to (8.center);
		\draw [bend left] (1) to (20.center);
		\draw [style=hadamard edge] (42) to (36);
		\draw [style=hadamard edge] (42) to (39);
		\draw [style=hadamard edge] (42) to (41);
		\draw (42) to (6);
		\draw [style=hadamard edge] (43) to (37);
		\draw [style=hadamard edge] (43) to (39);
		\draw [style=hadamard edge] (43) to (38);
		\draw (43) to (5);
	\end{pgfonlayer}
\end{tikzpicture}

%% file: figures/gen-pivot-mul/2.tikz
\begin{tikzpicture}
	\begin{pgfonlayer}{nodelayer}
		\node [style=Z dot] (0) at (-3.5, 1) {};
		\node [style=Z dot] (1) at (3.5, 1) {};
		\node [style=Z phase dot] (2) at (-1.25, -1) {$\beta_1 + \pi$};
		\node [style=Z phase dot] (3) at (1.25, -1.75) {$\beta_2 + \pi$};
		\node [style=Z phase dot] (4) at (-4.75, -2.25) {$\beta_3$};
		\node [style=Z phase dot] (5) at (-2.5, -4.5) {$\beta_4$};
		\node [style=Z phase dot] (6) at (4.75, -2.25) {$\beta_6$};
		\node [style=Z phase dot] (7) at (-3.5, 2.25) {$\alpha_2$};
		\node [style=Z phase dot] (8) at (3.5, 2.25) {$\alpha_1$};
		\node [style=none] (9) at (-0.175, -1.55) {...};
		\node [style=none] (10) at (3.5, -3) {...};
		\node [style=none] (11) at (-3.5, -3) {...};
		\node [style=none] (12) at (0, 2) {...};
		\node [style=Z phase dot] (13) at (2.5, -4.5) {$\beta_5$};
		\node [style=Z dot] (14) at (-3.5, 3.5) {};
		\node [style=none] (15) at (-3, 4.75) {};
		\node [style=none] (16) at (-4, 4.75) {};
		\node [style=none] (17) at (-3.475, 4.5) {...};
		\node [style=Z dot] (18) at (3.5, 3.5) {};
		\node [style=none] (19) at (4, 4.75) {};
		\node [style=none] (20) at (3, 4.75) {};
		\node [style=none] (21) at (3.525, 4.5) {...};
		\node [style=Z dot] (22) at (-1.25, 3.25) {};
		\node [style=none] (23) at (-1.75, 4.75) {};
		\node [style=none] (24) at (-0.75, 4.75) {};
		\node [style=none] (25) at (-1.25, 4.5) {...};
		\node [style=Z dot] (26) at (1.25, 3.25) {};
		\node [style=none] (27) at (0.75, 4.75) {};
		\node [style=none] (28) at (1.75, 4.75) {};
		\node [style=none] (29) at (1.25, 4.5) {...};
		\node [style=Z dot] (30) at (5.5, -1.25) {};
		\node [style=none] (31) at (5.75, 0) {};
		\node [style=none] (32) at (6.75, 0) {};
		\node [style=none] (33) at (6.2, -0.275) {...};
		\node [style=Z dot] (34) at (-5.5, -1.25) {};
		\node [style=none] (35) at (-5.75, 0) {};
		\node [style=none] (36) at (-6.75, 0) {};
		\node [style=none] (37) at (-6.125, -0.25) {...};
		\node [style=Z dot] (38) at (3.5, -4) {};
		\node [style=none] (39) at (4.25, -3.25) {};
		\node [style=none] (40) at (5.5, -3.25) {};
		\node [style=none] (41) at (4.75, -3.275) {...};
		\node [style=Z dot] (42) at (-3.75, -4) {};
		\node [style=none] (43) at (-4.5, -3.25) {};
		\node [style=none] (44) at (-5.75, -3.25) {};
		\node [style=none] (45) at (-5, -3.3) {...};
	\end{pgfonlayer}
	\begin{pgfonlayer}{edgelayer}
		\draw [style=hadamard edge] (0) to (2);
		\draw [style=hadamard edge] (0) to (3);
		\draw [style=hadamard edge] (3) to (1);
		\draw [style=hadamard edge] (2) to (1);
		\draw [style=hadamard edge] (0) to (4);
		\draw [style=hadamard edge] (0) to (5);
		\draw [style=hadamard edge] (1) to (6);
		\draw [style=hadamard edge] (0) to (1);
		\draw [style=hadamard edge] (4) to (2);
		\draw [style=hadamard edge] (5) to (3);
		\draw [style=hadamard edge] (4) to (6);
		\draw [style=hadamard edge] (2) to (6);
		\draw [style=hadamard edge] (1) to (13);
		\draw [style=hadamard edge] (5) to (13);
		\draw [style=hadamard edge] (3) to (13);
		\draw [style=hadamard edge] (0) to (7);
		\draw [style=hadamard edge] (1) to (8);
		\draw [bend right] (14) to (15.center);
		\draw [bend left] (14) to (16.center);
		\draw (7) to (14);
		\draw [bend right] (18) to (19.center);
		\draw [bend left] (18) to (20.center);
		\draw (8) to (18);
		\draw [bend left=15, looseness=0.75] (22) to (23.center);
		\draw [bend right=15, looseness=0.75] (22) to (24.center);
		\draw (2) to (22);
		\draw [bend left=15, looseness=0.75] (26) to (27.center);
		\draw [bend right=15, looseness=0.75] (26) to (28.center);
		\draw (26) to (3);
		\draw [bend left=15, looseness=0.75] (30) to (31.center);
		\draw [bend right, looseness=0.75] (30) to (32.center);
		\draw (6) to (30);
		\draw [bend right=15, looseness=0.75] (34) to (35.center);
		\draw [bend left, looseness=0.75] (34) to (36.center);
		\draw (4) to (34);
		\draw [bend left=15] (38) to (39.center);
		\draw [bend right=15] (38) to (40.center);
		\draw (13) to (38);
		\draw [bend right=15] (42) to (43.center);
		\draw [bend left=15] (42) to (44.center);
		\draw (5) to (42);
	\end{pgfonlayer}
\end{tikzpicture}

%% file: figures/gen-pivot-mul/3.tikz
\begin{tikzpicture}
	\begin{pgfonlayer}{nodelayer}
		\node [style=Z dot] (0) at (-3.5, 1) {};
		\node [style=Z dot] (1) at (3.5, 1) {};
		\node [style=Z phase dot] (2) at (-1.25, -1) {$\beta_1 + \pi$};
		\node [style=Z phase dot] (3) at (1.25, -1.75) {$\beta_2 + \pi$};
		\node [style=Z phase dot] (4) at (-4.75, -2.25) {$\beta_3$};
		\node [style=Z phase dot] (5) at (-2.5, -4.5) {$\beta_4$};
		\node [style=Z phase dot] (6) at (4.75, -2.25) {$\beta_6$};
		\node [style=Z phase dot] (7) at (-3.5, 2.25) {$\alpha_2$};
		\node [style=Z phase dot] (8) at (3.5, 2.25) {$\alpha_1$};
		\node [style=none] (9) at (-0.175, -1.55) {...};
		\node [style=none] (10) at (3.5, -3) {...};
		\node [style=none] (11) at (-3.5, -3) {...};
		\node [style=none] (12) at (0, 1.75) {...};
		\node [style=Z phase dot] (13) at (2.5, -4.5) {$\beta_5$};
		\node [style=none] (14) at (-3, 4) {};
		\node [style=none] (15) at (-4, 4) {};
		\node [style=none] (16) at (-3.475, 3.5) {...};
		\node [style=none] (17) at (4, 4) {};
		\node [style=none] (18) at (3, 4) {};
		\node [style=none] (19) at (3.525, 3.5) {...};
		\node [style=none] (20) at (-1.75, 4) {};
		\node [style=none] (21) at (-0.75, 4) {};
		\node [style=none] (22) at (-1.25, 3.25) {...};
		\node [style=none] (23) at (0.75, 4) {};
		\node [style=none] (24) at (1.75, 4) {};
		\node [style=none] (25) at (1.25, 3.25) {...};
		\node [style=none] (26) at (5, -1) {};
		\node [style=none] (27) at (6, -1) {};
		\node [style=none] (28) at (5.375, -1.275) {...};
		\node [style=none] (29) at (-5, -1) {};
		\node [style=none] (30) at (-6, -1) {};
		\node [style=none] (31) at (-5.35, -1.25) {...};
		\node [style=none] (32) at (3.25, -3.75) {};
		\node [style=none] (33) at (4.5, -3.75) {};
		\node [style=none] (34) at (3.775, -3.775) {...};
		\node [style=none] (35) at (-3.25, -3.75) {};
		\node [style=none] (36) at (-4.5, -3.75) {};
		\node [style=none] (37) at (-3.775, -3.8) {...};
	\end{pgfonlayer}
	\begin{pgfonlayer}{edgelayer}
		\draw [style=hadamard edge] (0) to (2);
		\draw [style=hadamard edge] (0) to (3);
		\draw [style=hadamard edge] (3) to (1);
		\draw [style=hadamard edge] (2) to (1);
		\draw [style=hadamard edge] (0) to (4);
		\draw [style=hadamard edge] (0) to (5);
		\draw [style=hadamard edge] (1) to (6);
		\draw [style=hadamard edge] (0) to (1);
		\draw [style=hadamard edge] (4) to (2);
		\draw [style=hadamard edge] (5) to (3);
		\draw [style=hadamard edge] (4) to (6);
		\draw [style=hadamard edge] (2) to (6);
		\draw [style=hadamard edge] (1) to (13);
		\draw [style=hadamard edge] (5) to (13);
		\draw [style=hadamard edge] (3) to (13);
		\draw [style=hadamard edge] (0) to (7);
		\draw [style=hadamard edge] (1) to (8);
		\draw [bend right=15] (7) to (14.center);
		\draw [bend left=15] (7) to (15.center);
		\draw [bend right=15] (8) to (17.center);
		\draw [bend left=15] (8) to (18.center);
		\draw [bend right=15, looseness=0.50] (3) to (24.center);
		\draw [bend left=15, looseness=0.50] (3) to (23.center);
		\draw [bend right=15, looseness=0.50] (2) to (21.center);
		\draw [bend left=15, looseness=0.50] (2) to (20.center);
		\draw [bend left=15] (13) to (32.center);
		\draw [bend right=15] (13) to (33.center);
		\draw [bend right=15] (5) to (35.center);
		\draw [bend left=15] (5) to (36.center);
		\draw [bend left=15] (6) to (26.center);
		\draw [bend right=15] (6) to (27.center);
		\draw [bend right=15] (4) to (29.center);
		\draw [bend left=15] (4) to (30.center);
	\end{pgfonlayer}
\end{tikzpicture}

%% file: figures/ga.tikz
\begin{tikzpicture}[FlowChart,
    node distance = 5mm and 7mm,
    start chain = A going below]
  \node (init) [startstop] {Initial population};                 
  \node (fit) [process]   {Fitness evaluation};        
  \node (sel) [process]   {Selection};        
  \node (gop) [process]   {Mutation \& crossover};   
  \node (ret) [startstop] {Return mutant with best fitness};                    
  \draw [arrow] (gop.east) node[right=15pt] {$\times n_{gens}$} -- + (1.0, 0) |- (fit);
  \draw [decorate,decoration={brace,mirror,amplitude=10pt, raise=5pt},xshift=-0.4pt,yshift=0pt]
  (fit.north west) -- (gop.south west) node [black,midway,xshift=-1.5cm]
        {\footnotesize Generation};
\end{tikzpicture}

%% file: chapters/results.tex
\chapter[Results]{Results} \label{ch:results}

Here we present the results of our optimization strategy.
Before reporting on the performance of our strategy, we first conduct preliminary analyses to refine our method.
All random circuits are generated using the \codeword{CNOT_HAD_PHASE_circuit} method in PyZX which constructs a circuit consisting of CNOT, HAD, and phase gates.
Default parameters are used for the probability of each gate type.


\section{Refinements}

General opportunities for refinement are the following:
\begin{itemize}
\item
  Optimization function
\item
  Probability of applying Equation \ref{eq:gen-io-lc} vs. Equation \ref{eq:gen-io-pivot}
\item
  Sampling of subjects (i.e., spiders or pairs of connected spiders) for congruence application
\end{itemize}
For these analyses, we only use SA as it is more computationally efficient and we expect the behavior of these refinements to generalize to GA.

We also conduct procedure-specific refinements.
We analyze how varying the number of iterations or mutants and generations affects optimization using SA and GA, respectively.

\subsection*{Optimization Function}

First, we test if any property of a ZX-diagram could serve as a reliable proxy for the complexity of its associated circuit.
To do so, we first generated 500 random circuits (10 qubits, 100 gates) and immediately converted them to ZX-diagrams.
Given this library of ZX-diagrams, we then measured the Pearson correlation coefficient between the complexity of the extracted circuit and each ZX-diagrammatic property discussed in Section \ref{sec:obj-funcs}.
The following table summarizes these correlations:
\begin{center}
\begin{tabular}[]{@{}l>{\centering\arraybackslash}p{1.5cm}>{\centering\arraybackslash}p{2cm}@{}}
\toprule
                    & $r$ & p-value \\ \midrule
\# Edges  & 0.097        & 0.029            \\
Centrality & 0.090        & 0.045            \\
Density    & -0.096       & 0.032            \\ \bottomrule
\end{tabular}
\end{center}
Pearson's correlation coefficient is denoted $r$ and the 2-tailed p-value is provided.
From these data, it is clear that no identified property of a ZX-diagram can serve as a reliable proxy for the complexity of its associated circuit.
Therefore, the default scoring method discussed in Section \ref{sec:strategy} that relies on extraction is used for the remainder of our analyses.

\begin{figure}[t]
\centering
\includegraphics[width=13cm]{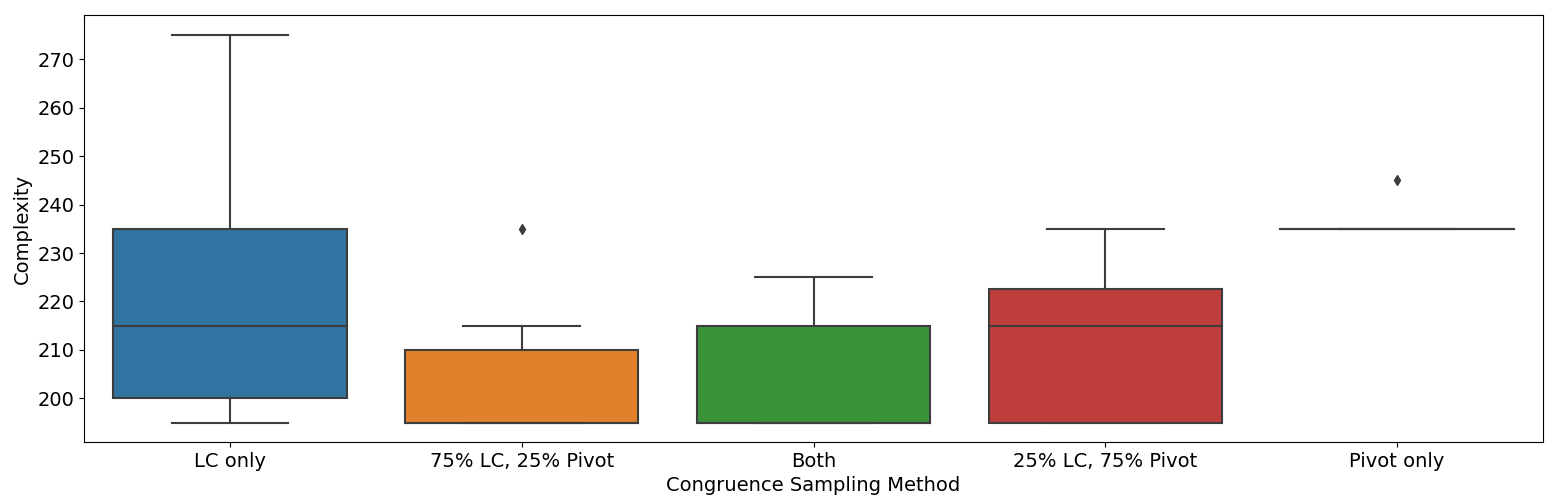}
\caption{
  A representative comparison of congruence sampling methods for a 5 qubit, 50 gate circuit.
  All three sampling methods that include both LC and pivoting achieve the best simplification in the alotted number of steps.
  When using only local complementation (i.e., Equation \ref{eq:gen-io-lc}), the same best-case complexity is achieved over the 10 trials but the average complexity is higher.
  Alternatively, using only pivoting does not achieve the same reduced circuit.
}
\label{fig:cong-sampling}
\end{figure}

\subsection*{Congruence Sampling}

By default, we apply Equations \ref{eq:gen-io-lc} and \ref{eq:gen-io-pivot} with equal probability ($p_{LC} = p_{pivot} = 0.5$).
However, it may be the case that one congruence should be chosen more frequently than the other.
To evaluate this, we generated 30 random circuits (5 qubits, 50 gates) and repeated search with a range of $(p_{LC}, p_{pivot})$ pairs.
For a given circuit, search was performed 10 times for each pair and the complexities of the optimized circuits were plotted according to congruence sampling probabilities.
In all cases, the sampling methods that include both LC and pivoting yielded the lowest average complexity across the 10 trials as well as the least complex circuit overall.
There were no significant differences between the three combined sampling methods (i.e., 50/50, 25/75, and 75/25), and the LC- ($p_{LC} = 1.0$, $p_{pivot} = 0.0$) or pivot-only ($p_{LC} = 0.0$, $p_{pivot} = 1.0$) approaches sometimes matched, but never outperformed these combined methods.
Most notably, the least complex circuit identified using LC-only matched the minimum complexity 73.3\% of the time while that identified using pivot-only was more complex than the best-case 86.7\% of the time.
A representative comparison for one random circuit is shown in Figure \ref{fig:cong-sampling}.

The observation that LC-only typically outperforms pivot-only is intuitive as one pivot is equivalent to three local complementations and therefore LC-only enables a more fine-grained search.
While LC-only likely converges to the combined sampling methods in the limit, we retain pivoting in the action set as it appears to require a fewer number of iterations at no cost.
In the remainder of our analyses, the default uniform sampling of Equations \ref{eq:gen-io-lc} and \ref{eq:gen-io-pivot} (i.e., $p_{LC} = p_{pivot} = 0.5$) is used.

\begin{figure}[t]
\centering
\includegraphics[width=13cm]{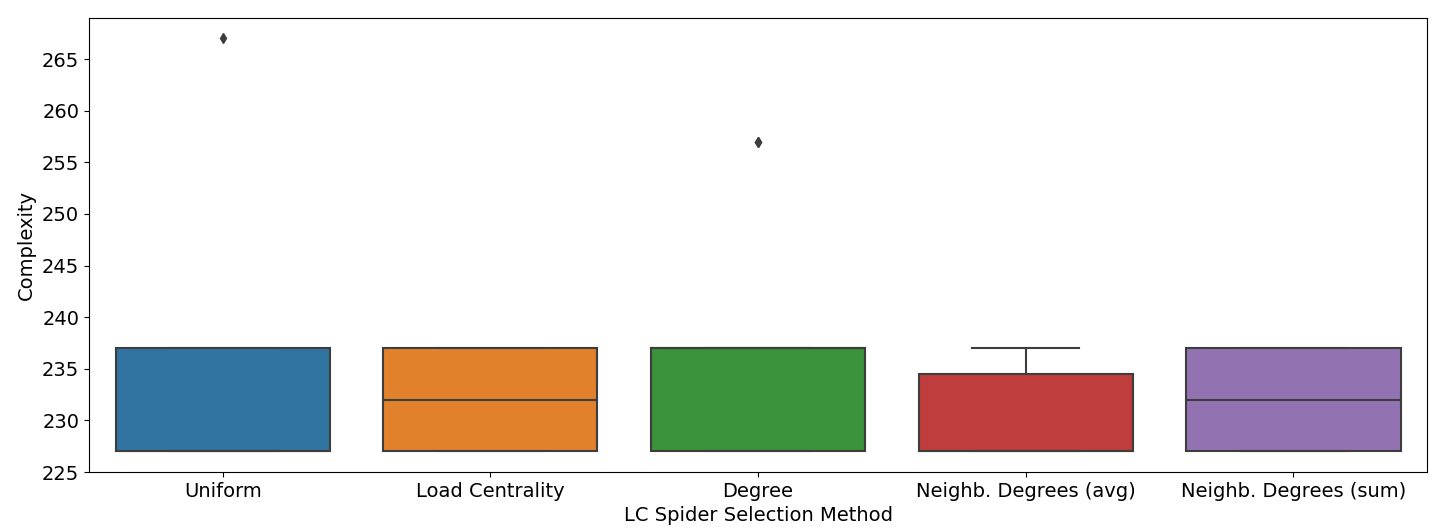}
\caption{
  A representative example of optimization of a 5 qubit, 50 gate circuit with different metrics of sampling spiders for application of Equation \ref{eq:gen-io-lc}.
  Optimization was performed 10 times for each sampling method using SA with $p_{LC} = 1.0$ and $p_{pivot} = 0.0$.
  No significant differences are observed in performance across the different weightings.
}
\label{fig:subj-sampling}
\end{figure}

\subsection*{Congruence Subject Sampling}

Similar to the probabilistic sampling of congruences to apply throughout search, we can experiment with methods of sampling subjects (i.e., spiders or pairs of connected spiders) for congruence application.
By default, we sample from the set of all eligible subjects uniformly.
However, we could alternatively weight this sampling in a way that improves search.
Candidate metrics for Equations \ref{eq:gen-io-lc} and \ref{eq:gen-io-pivot} are discussed in Section \ref{sec:strategy}.

To test these alternative weightings, we employ a similar approach as for congruence sampling.
First, we restrict the action space to that congruence for which we are testing various weightings to isolate its effect.
For example, if we are testing various weighting metrics to select a spider for application of Equation \ref{eq:gen-io-lc}, we set $p_{LC} = 1.0$ and $p_{pivot} = 0.0$.
We then proceed as before, evaluating the performance of SA on random circuits across 10 trials for each candidate weighting.
We evaluate 20 random circuits (5 qubits, 50 gates) for each set of sampling procedures.

In both cases, no sampling method demonstrated consistent improvement over any other across the 20 trials.
One representative example of this comparison for sampling spiders for Equation \ref{eq:gen-io-lc} is shown in Figure \ref{fig:subj-sampling}.
Similar results were observed for sampling pairs of connected spiders for Equation \ref{eq:gen-io-pivot}.
So, for the remainder of our analyses, we sample subjects for congruence application uniformly.

\begin{figure}[t]
\centering
\includegraphics[width=13cm]{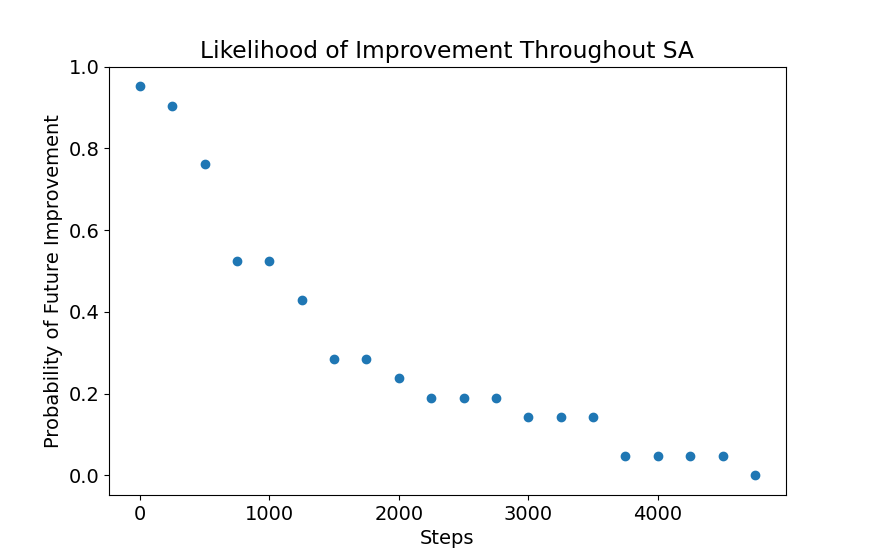}
\caption{
  The likelihood of obtaining a ZX-diagram with a simpler extracted circuit after a given iteration in SA.
  For example, after 2000 steps, there is a \textasciitilde 20\% chance of obtaining a ZX-diagram corresponding to an even simpler circuit as search progresses.
  These data were obtained using circuits with 4-16 qubits and 10-20 gates per qubit and do not necessarily generalize to larger circuits.
}
\label{fig:iter-likelihood}
\end{figure}

\subsection*{Number of SA Iterations}

The maximum number of SA steps $k_{max}$ should be high enough to permit the bulk of optimization while low enough to be computational feasible.
We determine the optimal $k_{max}$ by optimizing a set of random circuits and computing the probability that SA uncovers a complexity reduction after fixed intervals.
First, we generated a set of 21 random circuits with 4-16 qubits (intervals of 2) and 10-20 gates per qubit (intervals of 5).
For each circuit, we obtained a simplified ZX-diagram via \codeword{teleport_reduce} + \codeword{basic_optimization}.
We then optimized each simplified ZX-diagram using SA with $k_{max} = 5000$ and recorded the complexity of the best circuit every 250 steps.
Lastly, we computed the probability that a ZX-diagram whose extracted circuit is less complex would be found as search progressed for each 250-step interval.

The results of this analysis are shown in Figure \ref{fig:iter-likelihood}.
We see that most improvement occurs in the beginning of search and that the chance of finding a simpler circuit
plateaus after \textasciitilde 2500 steps.
For this reason, we fix $k_{max} = 2500$ for the remainder of our analyses unless otherwise noted.

\begin{figure}
\centering
\begin{subfigure}[t]{0.47\textwidth}
  \centering
  \includegraphics[width=\linewidth]{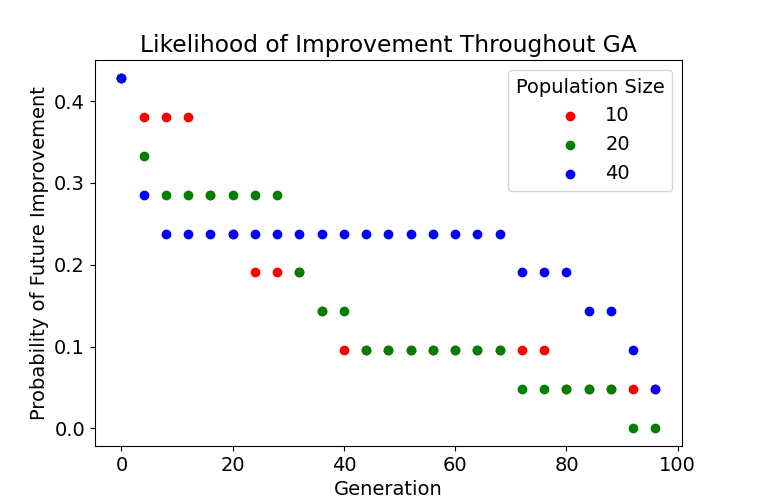}
  \caption{
    The likelihood of obtaining a simpler circuit after a given generation.
    As in Figure \ref{fig:iter-likelihood}, these data were obtained using circuits with 4-16 qubits and 10-20 gates per qubit.}
  \label{fig:ga-likelihood}
\end{subfigure}
\hfill
\begin{subfigure}[t]{0.47\textwidth}
  \centering
  \includegraphics[width=\linewidth]{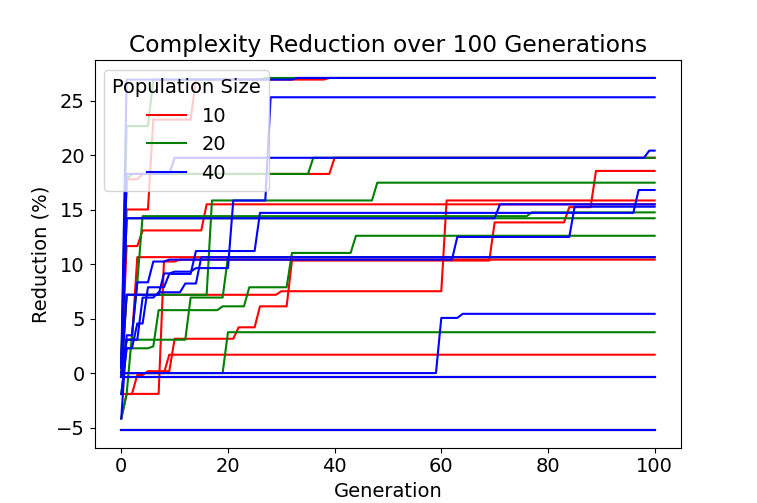}
  \caption{
    The reduction in complexities over time for 10 random 7 qubit, 100 gate circuits.
    Reduction in complexity is measured with respect to the simplified ZX-diagram rather than the original circuit.
  }
  \label{fig:7qb}
\end{subfigure}
\caption{
  GA parameter tuning
}
\label{fig:ga-params}
\end{figure}

\subsection*{GA Parameters}

Similarly, when searching this space with GA, we want to evolve a large enough population for sufficiently many generations while incurring minimal computational cost.
To determine the appropriate population size and number of generations, we employ a similar approach as for determining $k_{max}$.
First, we generated a set of 21 simplified ZX-diagrams in the same fashion.
We then evolved each ZX-diagram for 100 generations with a population of 10, 20, or 40 mutants.
Lastly, for each population size, we computed the probability that a ZX-diagram corresponding to a simpler circuit (via extraction) would be found as search progressed for each 2-generation interval.
These probabilities for each population size are shown in Figure \ref{fig:ga-likelihood}.
All three population sizes exhibit similar behavior, with the largest population ($n_{mutants} = 40$) experiencing more improvement in later generations.

We also sought to understand both the rate of reduction (with respect to the simplified ZX-diagram) over time and differences in total reduction between population sizes.
We plotted the complexity reduction over time for 10 random circuits with 7 qubits and 100 gates for each population size.
These reductions (one line per circuit for each population size) are shown in Figure \ref{fig:7qb}.
It is immediately clear that most optimization occurs in the first few generations which is supported by the low likelihood of improvement in early generations shown in Figure \ref{fig:ga-likelihood}.
Additionally, there is no reliable benefit provided by a larger population size.
We fix $n_{mutants} = 20$ and $n_{gens} = 40$ for the remainder of our analyses unless otherwise stated.

\begin{table}[t]
  \centering
\begin{tabular}{@{}ccccccc@{}}
\toprule
\multirow{3}{*}{\textbf{Qubits}}           & \multicolumn{6}{c}{\textbf{Complexity Reduction (\%)}}                                                      \\ \cmidrule(l){2-7}
                                           & \multirow{2}{*}{TR} & \multirow{2}{*}{FR} & \multicolumn{2}{c}{SA} & \multicolumn{2}{c}{GA} \\ \cmidrule(l){4-7}
                                           &                     &                     & TR seed    & FR seed   & TR seed    & FR seed   \\ \midrule
4                                          & 32.0                & 40.1                & 57.8       & 57.3      & 37.0       & 39.3      \\
6                                          & 10.3                & 9.4                 & 33.0       & 31.4      & 6.2        & 5.8       \\
8                                          & 2.1                 & -27.8               & 5.1        & 4.4       & -22.9      & -19.7     \\
10                                         & 5.1                 & -40.3               & -15.5      & -17.2     & -44.1      & -39.6     \\
12                                         & 1.2                 & -65.0               & -32.6      & -32.3     & -60.8      & -65.8     \\
14                                         & 1.4                 & -77.6               & -47.4      & -48.2     & -78.3      & -76.8     \\ \midrule
\multicolumn{1}{l}{\textbf{Avg. Time (s)}} & 0.03                & 0.04                & 123.1      & 122.4     & 35.8       & 33.4      \\ \bottomrule
\end{tabular}
\caption{\label{tab:init-results}
  A comparison of our optimization strategy with off-the-shelf PyZX methods using default parameters for varying circuit sizes (15 gates per qubit).
  Complexity reduction is measured with respect to the original circuit.
  TR and FR denote the two standard optimization pipelines in PyZX based on {\color{gray}\texttt{teleport\_reduce}} and {\color{gray}\texttt{full\_reduce}}, respectively.
  TR seed and FR seed denote the two methods of obtaining an initial simplified ZX-diagram (see Figure \ref{fig:primary}).
}
\end{table}

\begin{table}[t]
  \scriptsize
  \centering
  \subfloat[15 gates per qubit]{
  \begin{tabular}{@{}ccccccc@{}}
  \toprule
  \multirow{2}{*}{\textbf{Qubits}} & \multicolumn{6}{c}{\textbf{Complexity Reduction (\%)}} \\ \cmidrule(l){2-7}
                                   & TR     & FR    & SA    & GA    & qiskit    & tket    \\ \midrule
  4                                & 32.0      & 40.1     & 58.6     & 45.1     & 24.7         & 47.0         \\
  6                                & 10.3      & 9.4     & 34.2     & 7.4     & 14.7         & 28.7         \\
  8                                & 2.1      & -27.8     & 7.6     & -22.9     & 9.2         & 20.1         \\
  10                               & 5.1      & -40.3     & -11.9     & -43.3     & 9.0         & 15.0         \\
  12                               & 1.2      & -65.0     & -30.1     & -66.0    & 6.4         & 12.0         \\
  14                               & 1.4      & -77.6     & -45.0     & -79.6     & 5.7         & 10.4         \\
  \bottomrule
  \end{tabular}}
  \quad
  \subfloat[40 gates per qubit]{
  \begin{tabular}{@{}ccccccc@{}}
  \toprule
  \multirow{2}{*}{\textbf{Qubits}} & \multicolumn{6}{c}{\textbf{Complexity Reduction (\%)}} \\ \cmidrule(l){2-7}
                                   & TR     & FR    & SA    & GA    & qiskit    & tket    \\ \midrule
  4                                & 41.1      & 61.3     & 70.5     & 63.0     & 22.6         & 49.1         \\
  6                                & 20.2      & 38.4     & 50.1     & 38.3     & 14.0         & 28.9         \\
  8                                & 12.1      & 11.0     & 26.3     & 10.9     & 11.3         & 22.2         \\
  10                               & 9.5      & -6.8     & 10.4     & -6.7     & 9.9         & 19.0         \\
  12                               & 8.0      & -28.5     & -10.6     & -27.0     & 8.5         & 16.3         \\
  14                               & 4.2      & -46.3     & -29.5     & -45.6     & 6.8         & 12.9         \\
  \bottomrule
  \end{tabular}}
  \caption{\label{tab:compare-results}
    A comparison of our optimization strategy with off-the-shelf methods using parameters that lend to increased runtime and performance.
    The optimization procedures in both {\color{gray}\texttt{qiskit}} and {\color{gray}\texttt{tket}} execute in $<1\text{s}$.
  }
\end{table}

\section{Performance}

After refining the parameters of our method, we sought to understand our method's ability to optimize a given input circuit.
For $4, 6, \dots 14$ qubits, we optimized the simplified ZX-diagrams of 10 randomly generated circuits with 15 gates per qubit.
We recorded the average complexity reduction with respect to the original circuit (rather than the simplified ZX-diagram) as well as the execution time.
We conducted this analysis for six optimization methods -- both off-the-shelf optimization pipelines in PyZX, and all four combinations of the two search procedures and the two methods for obtaining an initial simplified ZX-diagram.
These data are shown in Table \ref{tab:init-results}.

Immediately we see that our strategy outperforms existing methods for low qubit numbers ($<10$).
For circuits with $\geq 10$ qubits, our method produces a circuit that is more complex than the original circuit.
For these larger circuits, our method performs worse than the \codeword{teleport_reduce} pipeline, which guarantees to not introduce complexity given its lack of circuit extraction, but introduces less additional complexity than the \codeword{full_reduce} pipeline.
Furthermore, neither method of obtaining a simplified ZX-diagram to seed search appears to provide benefit over the other.
However, SA consistently outperforms GA.

\begin{table}[t]
  \footnotesize
  \centering
\begin{tabular}{@{}ccccccccccccc@{}}
\toprule
 &
   &
   &
   &
  \multicolumn{4}{c}{\textbf{Comp. Reduction (\%)}} &
  \multicolumn{1}{l}{} &
  \multicolumn{4}{c}{\textbf{2-qubit Count}} \\ \cmidrule(lr){5-8} \cmidrule(l){10-13}
\textbf{Circuit} &
  \textbf{n} &
  \textbf{Gates} &
  \textbf{2-qb} &
PyZX &
  SA &
  qiskit &
  tket &
   &
  PyZX &
  SA &
  qiskit &
  tket \\ \midrule
barenco-tof-3                & 5  & 58   & 24  & 0.7  & 29.9  & 1.8  & 2.2  &  & 24  & 17  & 24  & 24  \\
barenco-tof-4                & 7  & 114  & 48  & 5.5  & 32.1  & 1.8  & 2.4  &  & 47  & 33  & 48  & 48  \\
barenco-tof-5                & 9  & 170  & 72  & 1.0  & 32.9  & 1.8  & 2.5  &  & 72  & 49  & 72  & 72  \\
$\text{gf}(2^4)\text{-mult}$ & 12 & 243  & 99  & 3.1  & -96.6 & 3.3  & 3.3  &  & 99  & 215 & 99  & 99  \\
grover-5                     & 9  & 831  & 288 & 4.3  & 19.5  & 4.7  & 6.6 &  & 288 & 248 & 288 & 288 \\
$\text{hwb}_6$               & 7  & 259  & 116 & 5.3  & 5.7   & 1.8  & 6.5  &  & 111 & 112 & 116 & 111 \\
$\text{mod5}_4$              & 5  & 63   & 28  & 14.3 & 52.4  & 2.2  & 2.2  &  & 26  & 14  & 28  & 28  \\
mod-mult-55                  & 9  & 119  & 48  & 2.9  & -16.0 & 1.6  & 3.6 &  & 48  & 59  & 48  & 48  \\
mod-red-21                   & 11 & 278  & 105 & 2.7  & 7.3   & 3.4  & 4.3 &  & 105 & 103 & 105 & 105 \\
$\text{nth-prime}_6$         & 9  & 1241 & 502 & 5.8  & 23.4  & 3.5  & 5.4  &  & 502 & 401 & 498 & 493 \\
$\text{rc-adder}_6$          & 14 & 200  & 93  & 13.2 & 31.4  & 11.7 & 14.3  &  & 81  & 64  & 83  & 81  \\
$\text{tof}_3$               & 5  & 45   & 18  & 0.5  & 22.2  & 1.9  & 2.4  &  & 18  & 14  & 18  & 18  \\
$\text{tof}_4$               & 7  & 75   & 30  & 0.6  & 26.7  & 2.0  & 2.6 &  & 30  & 22  & 30  & 30  \\
$\text{tof}_5$               & 9  & 105  & 42  & 0.6  & 22.2  & 2.1  & 2.7  &  & 42  & 33  & 42  & 42  \\
$\text{vbe-adder}_3$         & 10 & 150  & 70  & 16.7 & 44.7  & 17.4 & 17.5  &  & 58  & 39  & 58  & 58  \\ \bottomrule
\end{tabular}
\caption{\label{tab:bench}
  An evaluation of our method on the benchmark circuits from \cite{kissinger2019reducing} with at most 14 qubits.
  The number of qubits of the input circuit is denoted $n$.
  We abbreviate the number of 2-qubit gates in the input circuit as 2-qb.
  The PyZX column corresponds to whichever of PyZX's two standard optimization pipelines (i.e., {\color{gray}\texttt{full\_reduce}} or {\color{gray}\texttt{teleport\_reduce}} followed by circuit-level optimizations) resulted in the best reduction for a given circuit.
}
\end{table}

Not shown in Table \ref{tab:init-results} is that in all optimization conducted via our method, the overall number of single-qubit gates never decreased from the initial simplified ZX-diagram.
This means that all reduction in complexity achieved via our method is a consequence of reducing the number of 2-qubit gates.
Notably, the T-count always remained unchanged.
We expand on this in our analysis of (non-random) benchmark circuits.

We then sought to both stress the performance of our method and place it in the broader landscape of circuit optimization techniques.
We conducted the same analysis with the following changes:
\begin{enumerate}
\item
  We only used \codeword{teleport_reduce} + \codeword{basic_optimization} to obtain a simplified ZX-diagram as there are no significant performance differences between the two methods.
\item
  We adjusted search parameters in a way that increases execution time but may lend to better performance.
  For SA, we used three random restarts rather than a single instantiation.
  For GA, we increased $n_{gens}$ to 75.
\item
  We included the standard optimization procedures in IBM's \codeword{qiskit}~\cite{Qiskit} and Cambridge Quantum Computing's \codeword{tket}~\cite{sivarajah2020t} libraries as baselines.
\end{enumerate}
We repeated this analysis on circuits with both 15 and 40 gates per qubit.

The results of this analysis are shown in Table \ref{tab:compare-results}.
First, we observe minor increases in the performance of annealing from the two additional random restarts.
Similarly, the additional generations in the evolutionary approach result in substantial improvements for the 4-qubit circuits but minor differences at other sizes;
SA continues to consistently outperform GA.
Secondly, while we again observe a decrease in performance with the number of qubits, our method appears to scale with the number of total gates in the input circuit.
For 10-qubit circuits with 400 gates, annealing yields a 10\% reduction in complexity compared to the increase in complexity at 15 gates per qubit.
This result aligns with the general trend that the circuits with 40 gates per qubit appear to be more reducible than those with fewer gates.
Lastly, our method outperforms both additional off-the-shelf methods at low numbers of qubits.

Lastly, we sought to evaluate our method on a non-random class of circuits.
We tested our approach alongside off-the-shelf methods on all benchmark circuits from \cite{kissinger2019reducing} with at most 14 qubits.
We only used SA given its continued benefit over GA and conducted three random initializations for each circuit.

Table \ref{tab:bench} provides a summary of these analyses.
Excitingly, our method outperforms existing libraries in 86.7\% of cases (all but two).
Most notably, our method performs far better on benchmark circuits with $>8$ qubits than on randomly generated circuits of similar sizes.
For example, our method reduces a 14-qubit circuit (i.e., $\text{rc-adder}_6$) by 31.4\%.
Furthermore, in all cases that annealing reduced circuit complexity, we observe significant reductions in 2-qubit count yet the number of single-qubit gates is never lower than the circuit extracted from the simplified ZX-diagram used to seed search (not shown).
This provides further evidence that our simplification procedure is only effective in reducing the 2-qubit count and at best does not increase the number of single-qubit gates.
We leave proving such a property to future work.


%% file: chapters/discuss-conc.tex
\chapter[Discussion and Conclusion]{Discussion and Conclusion} \label{ch:discuss-conc}

We first tested our method on randomly generated circuits.
For both SA and GA, we observed a clear decrease in performance with the number of qubits.
Performance increased with higher-depth circuits, though the trend remained.
One explanation for this trend is that gate counts were increased linearly with the number of qubits which corresponds to a sub-linear increase in connectivity between qubits.
In future work, it would interesting to explore how performance scales if depth is increased polynomially with the number of qubits.
Alternatively, it may be that circuit extraction is less efficient for larger circuits;
addressing this issue would require an alternative extraction procedure.
Additionally, SA always outperformed GA and the method of obtaining a simplified ZX-diagram to seed search did not seem to affect performance.

We then tested how SA performed on a suite of benchmark circuits with $\leq 14$ qubits.
Interestingly, the magnitude of the scaling issues encountered with random circuits was severely reduced;
we drastically outperformed off-the-shelf methods for all but two benchmark circuits.
While the two circuits for which our method did not perform well have $>10$ qubits, SA succeeded in reducing the complexity of multiple benchmark circuits with $>8$ qubits including a 14-qubit circuit by 31.4\%.
This is a promising result and suggests that our procedure may perform differently on varying classes of circuits.
In future work, we hope to test our method on well-defined circuit classes (e.g., quantum chemistry circuits) and to better understand the class of circuits for which our method performs well.

In this work, we applied search over local variants of a ZX-diagram that had been simplified to fixpoint.
However, it may be that fully simplifying a ZX-diagram may induce complexity upon extraction (e.g., by eliminating spiders at the cost of high connectivity).
We plan to test this by seeding search with a ZX-diagram that is not fully-simplified.
The most ambitious form of this variant would be including congruences in a set of simplification rules to form a master action set over which search could be applied given the original circuit (i.e., no initial simplification of the ZX-diagram).
Alternatively, we could locally apply search during the extraction procedure itself to minimize the number of CNOTs introduced per edge.

Though a work in progress, this new approach to quantum circuit optimization shows promise given these preliminary results on both random and benchmark circuits.
There are many opportunities for improvement, from including more congruences or improving the extraction procedure to changing when search is applied altogether.
All of these potential paths forward demonstrate the power and flexibility of the ZX-calculus and its crucial role in reasoning about quantum processes.